\documentclass[12pt]{iopart}

\usepackage{graphics,epsfig}
\begin{document}

\title[Cold molecules formation by shaping with strong light the short-range interaction]{Cold molecules formation by shaping with light the short-range interaction between cold atoms: photoassociation with strong laser pulses}

\author{Mihaela Vatasescu}

\address{Institute of Space Sciences, MG-23, 77125 Bucharest-Magurele, Romania}
\ead{mihaela@venus.nipne.ro}
\begin{abstract}
The paper investigates cold molecules formation in the photoassociation of two cold atoms 
by a strong laser pulse applied at short interatomic distances, which lead to a molecular dynamics taking place in the light-induced (adiabatic) potentials. A two electronic states model in the cesium dimer is used to analyse the effects of this strong coupling regime and to show specific results: i) acceleration of the ground state population to the inner zone due to a non-impulsive regime of coupling at short and intermediate interatomic distances; ii) formation of cold molecules in strongly bound levels of the ground state, where the population at the end of the pulse is much bigger than the population photoassociated in bound levels of the excited state; iii)
the final momentum distribution of the ground state wavepacket keeping the signatures of the maxima in the initial wavefunction continuum. It is shown that the topology of the light-induced potentials plays an important role in dynamics.
\end{abstract}


\pacs{31.15.xv, 32.80.Qk, 33.15.Vb, 33.20.Tp, 34.50.Cx}
\submitto{\jpb}
\maketitle

\section{Introduction}

Photoassociation of cold atoms as a technique to form cold molecules has known new developments in the last years due to explorations using shaped laser pulses in order to control molecules formation:  enhancement in cold molecules production and attainment of deeply bound vibrational states are both desired. The road from cold atoms photoassociation using continuous lasers \cite{julienne06} which keep the selectivity of transitions to the use of ultrashort pulses with broad bandwidth brought the challenges of a new physics. Prospective experiments using shaped femtoseconds laser pulses to photoassociate ultracold atoms \cite{salzmann06,walmsley06} have shown the coherence of the process, but emphasized the difficulty to increase the number of created molecules. On the other hand, frequency-chirped light pulses in the nanosecond range were used to coherently control ultracold atomic Rb collisions showing the enhancement of short-range collisional flux \cite{gould06,gould07}. Other experimental developments trying to establish coherent control techniques for cold molecules formation explored multiphoton photoassociative ionization 
in a Rb magneto-optical trap combining femtoseconds and continuous lasers \cite{veshapidze07} and  optimal control of multiphoton ionization of Rb$_2$ molecules using femtosecond laser pulses in a closed feedback loop \cite{weise07}. Theoretical studies of pulsed photoassociation explored a variety of schemas to control cold molecular dynamics: with chirped pulses \cite{vala01,chirpPRA04,elianeepjd04,elihole07}, pump-dump schemes to stabilize the cold molecules \cite{kochluc06,kochRb06,jordidyn07,kallush07}, schemes using adiabatic passage \cite{shapiro07,shapiro08}.

The present paper prolonges previous works \cite{vatasescu01,vatasescu08}, the aim being to investigate theoretically the photoassociation of two cold atoms 
by strong laser pulses applied
 at small or intermediate interatomic distances. Such pulses have to be only
 ``moderately strong'' in order to avoid additional processes as
 ionization, and to act far from the atomic resonance
 (i.e. with a large red detuning, the colliding atoms being excited at 
small interatomic separations), to avoid the transfer of population
 to the continuum of the excited state. This regime of strong coupling and large detuning in cold atoms photoassociation can be used to address some specific interrogations: i) One interest is to explore if a strong pulse applied at small interatomic distances could be used to create strongly bound cold molecules, i.e. if such a pulse could accelerate efficiently the initial population, located at large interatomic distances in a cold collision, towards the inner region. ii) Secondly, a transition taking place at small or intermediate interatomic distances brings a regime necessarily ``seeing`` the nodal structure  of the initial ground state continuum, and then we expect such traces in our results. iii) We are interested to explore the effects produced by a rather strong
regime of coupling, generally not easy to be predicted. In \cite{vatasescu01} we have shown that for a strong coupling, characteristic times related to the adiabatic potentials become relevant in dynamics. The present work will explicitely explore the role played by the topology of the light-induced potentials in the dynamics and the photoassociation results.

Our analysis is pursued on the example of the $a^3\Sigma_{u}^{+}(6s,6s)$ $\to$ $1_g(6s,6p_{3/2})$ transition in Cs$_2$. The structure of the paper is the following: \Sref{sec:model} describes the time-dependent model
 and the time scales relevant in the photoassociation dynamics, as well as the initial wavefunction and the time evolution of the wavepackets during the pulse. In \Sref{sec:poptransfer} we analyse the population transfer during the pulse and  show the relevance of the light-induced potentials for the dynamics. In \Sref{sec:endpulse} are shown and interpreted the results at the end of the pulse: formation of strongly bound cold molecules in ground and excited electronic states (\Sref{sec:finpop}), acceleration of the ground state population to small interatomic distances (\Sref{sec:accground}), and the momentum structure of the final wavepacket in the ground state which reflects the maxima of the initial continuum wavefunction
(\Sref{sec:kfeatures}). An Appendix is connected to \Sref{sec:kfeatures}. \Sref{sec:conclu} contains comments and conclusions.

\section{Simulation of the photoassociation dynamics during the pulse}
\label{sec:model}

\begin{figure} 
\center
\includegraphics[width=0.6\columnwidth]{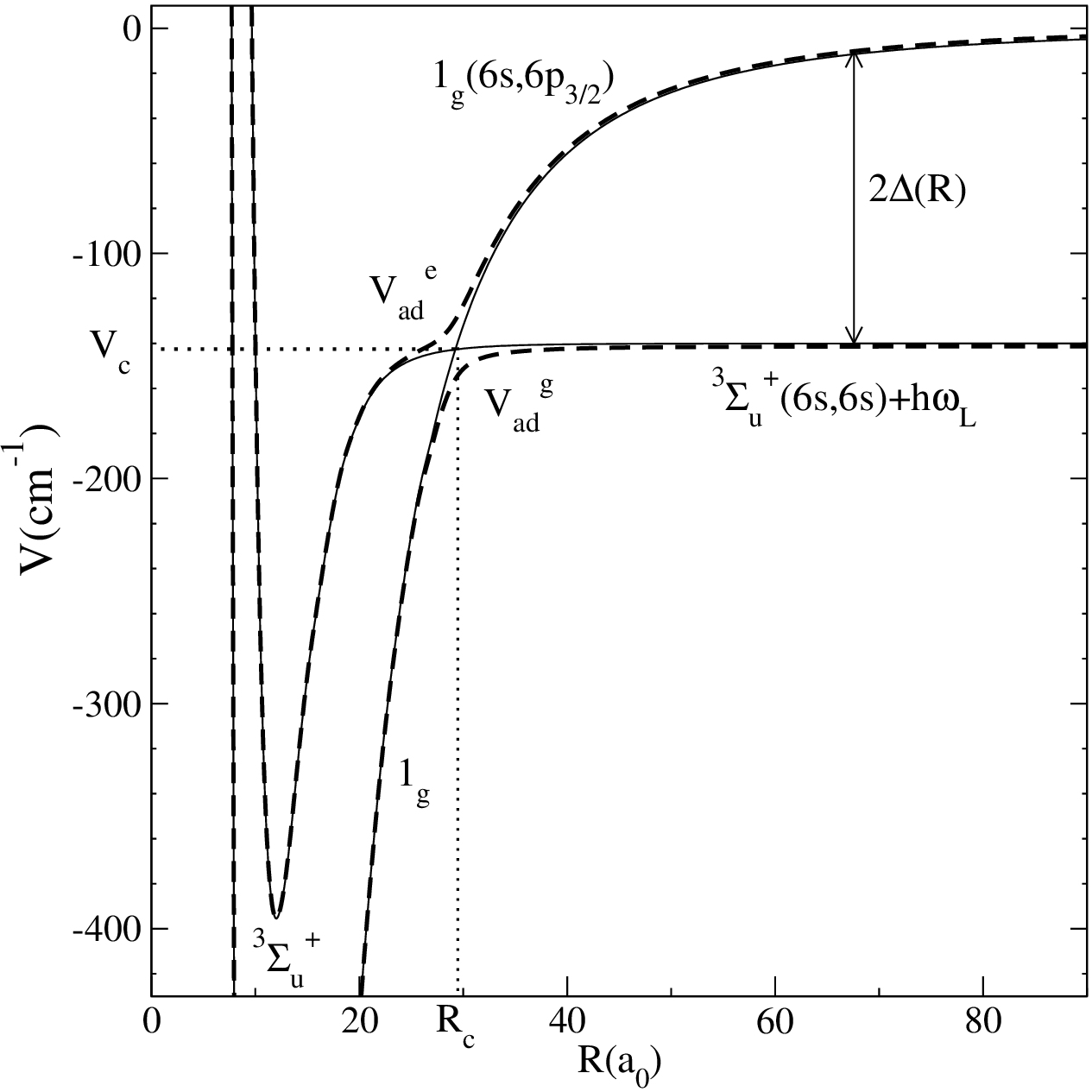}
          \caption{$a^3\Sigma_{u}^{+}(6s,6s)$ and $1_g(6s,6p_{3/2})$ electronic
          potentials of Cs$_2$ (full lines), dressed with the photon of energy
          $\hbar\omega_L$= $E_{6p_{3/2}}-E_{6s} - \hbar\Delta_L$ ($\hbar
          \Delta_L$=140 cm$^{-1}$) and
          crossing in $R_c=29.3 \ a_0$, $V_c$=$V_{1_g}(R_c)$=$V_{\Sigma}(R_c)$=
          -143 cm$^{-1}$. Dashed lines: the adiabatic potentials ($V_{ad}^e$, $V_{ad}^g$) obtained from the
          diagonalization of the 2x2 potential matrix 
          \eref{adiabmatrix} with the
          coupling  $W_L=13.17$ cm$^{-1}$. The energy origin is
          taken to be the dissociation limit $E_{6s+6p_{3/2}}=0$ of the
          $1_g(6s+6p_{3/2})$ potential.} 
\label{fig1_pot}
\end{figure}
The photoassociation reaction is between two cold
cesium atoms colliding in the ground state potential $g=a^3 \Sigma_u^+(6s,6s)$, at a temperature $T =$ 0.11 mK, which are excited  by a moderately strong laser pulse (with intensity I $\approx$ 43 MW/cm$^{2}$) to form a molecule in a superposition of vibrational levels $\{ v_e \}$ of the excited electronic potential $e=1_g(6s,6p_{3/2})$. Only the $s$ wave of the collision is considered. For a rotational quantum number $J=0$, the process can be schematized as:
\begin{eqnarray}
Cs(6s)+Cs(6s)+ \hbar\omega_L  \rightarrow Cs_{2}(1_g(6s+6p_{3/2}); \{ v_e \}, J=0)
\label{eq:photo}
\end{eqnarray}
 The pulse is red-detuned with $\hbar\Delta_L=140$ cm$^{-1}$ from the energy $\hbar\omega_{at}=
E_{6p_{3/2}}-E_{6s}$ of the D2 atomic transition ($\hbar\omega_L$=$\hbar \omega_{at}$- $\hbar \Delta_L$). The large detuning $\hbar\Delta_L$ determines a crossing 
of the field dressed diabatic potentials at the interatomic distance $R_c=29.3 \ a_0$, with 
$V_c$=$V_{1_g}(R_c)$=$V_{\Sigma}(R_c)$=-143 cm$^{-1}$ (\fref{fig1_pot}).

The $a^3\Sigma_{u}^{+}(6s,6s)$ and $1_g(6s,6p_{3/2})$ electronic potentials used in the present calculation (\fref{fig1_pot}) are built from quantum chemistry \cite{spies89} and asymptotic calculations \cite{marinescu95,marinescu96} and were described in a previous paper \cite{vatasescu01}. 

\subsection{Time-dependent model}
\label{sec:tdepmodel}

The dynamics of the photoassociation process is simulated by solving numerically the time-dependent Schr\"odinger equation associated with the radial
motion of the wavepackets $\Psi_{1_g}(R,t)$ and $\Psi_{\Sigma}(R,t)$ in the electronic channels $1_g$ and $a^3\Sigma_u^+$, coupled by an electric field with the amplitude
${\cal{E}}(t)={\cal {E}}_0 f(t) \cos \omega_Lt$. The equation can be written as \cite{vatasescu01}:
\begin{eqnarray}
\label{tschreq}
&&i\hbar\frac{\partial}{\partial t}\left(\begin{array}{c}
 \Psi_{1_g}(R,t)\\
\Psi_{\Sigma}(R,t)
 \end{array}\right)=\\
&&
\left(\begin{array}{lc}
 {\bf \hat T} + V_{1_g}(R)  &
W_L f(t) \\
 W_L f(t) &
 {\bf \hat T} + V_{\Sigma}(R)
 \end{array} \right)
 \left( \begin{array}{c}
 \Psi_{1_g}(R,t)\\
\Psi_{\Sigma}(R,t)
 \end{array} \right) \nonumber
 \end{eqnarray}

\Eref{tschreq} is obtained in the Born-Oppenheimer approximation for the diatomic molecule and using the rotating wave approximation with the frequency  $\omega_L/2\pi$.
The potentials $V_{1_g}(R)$ and $V_{\Sigma}(R)$ are the diabatic electronic
potentials crossing in $R_c$, represented in \fref{fig1_pot}.
${\bf \hat T}$ is the kinetic energy operator
and $W_L f(t)$ the coupling between the two channels, with $f(t)$ the temporal envelope of the pulse. $W_L= - \frac {1}{2}{\cal {E}}_0 D_{ge}^{\vec{e_L}}$, where
${\cal {E}}_0=\sqrt{\frac{2I}{c\epsilon_0}}$ is the field amplitude
(with $I$ the laser intensity), ${\vec{e_L}}$ the polarization, and $D_{ge}^{\vec{e_L}}$  the transition dipole moment between the ground and the excited molecular
electronic states. We neglect the R-dependence of the transition dipole moment, using the asymptotic value $D_{ge}^{\vec{e_L}}$ deduced from standard long-range calculations 
for a linear polarisation vector $\vec{e_L}$. This approximation remains good for the present calculation, as for distances around the crossing $R_c=29.3 \ a_0$, the dipole moment is closed to its asymptotic value, $D_{ge}^{\vec{e_L}}(R_c)$ $\approx$ 0.9 $D_{ge}^{\vec{e_L}}$ , and decreases slowly for smaller distances \cite{vatasescu99}.
 For a pulse intensity I $\approx$ 43 MW/cm$^{2}$, the coupling becomes W$_L$=13.17 cm$^{-1}$,  inducing a significant avoided crossing, as it can be seen in \fref{fig1_pot}, where the light-induced or adiabatic potentials $V_{ad}^e$, $V_{ad}^g$ are represented with dashed lines.

\begin{figure} 
\center
\includegraphics[width=0.4\columnwidth]{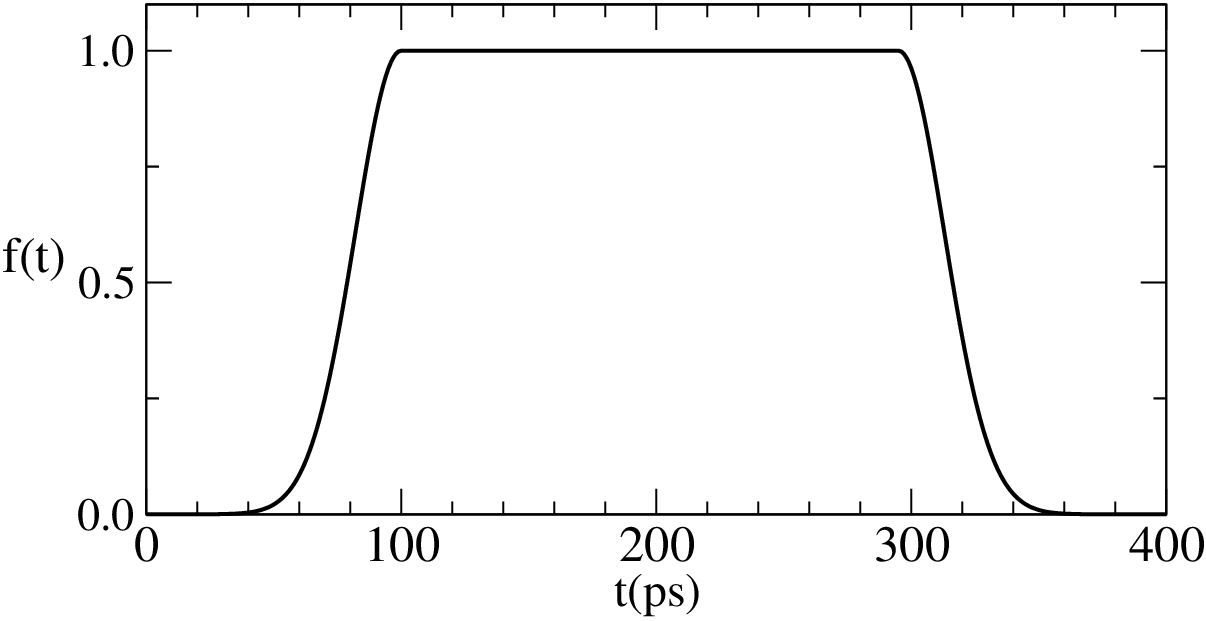}
          \caption{Temporal envelope f(t) of the photoassociating pulse.}
\label{fig2_pulse}
\end{figure}

The numerical calculations were made for a rectangular pulse with a duration of $\approx$ 300
ps, whose envelope  f(t) is  represented in \fref{fig2_pulse}.

The Schr\"odinger equation \eref{tschreq} is solved by propagating in time
 an initial wavefunction $\left(\begin{array}{c}
 0 \\
\Psi_{\Sigma}(R,0)
 \end{array}\right)$ on a spatial grid with the length $L_R = 760$ a$_0$. The time propagation uses the Chebychev expansion of the evolution
 operator  \cite{kosloff94,kosloff96} and the Mapped Sine Grid (MSG) method
 \cite{elianeepjd04, willner04}  to represent the radial dependence of the
 wavepackets.

The results extracted  from the dynamics are:
\begin{itemize}
\item the evolution of the wavepackets during the pulse, for the
two channels $g=a^3\Sigma_u^+$, $e=1_g$, in the position
representation, $\Psi_{\Sigma,1g}(R,t)$, and  momentum representation,
defined by the Fourier transforms $\Psi_{\Sigma,1g}(p,t)$;
\item the evolution of the population in each electronic state during the pulse.  At a given instant $t$, the population in one of the electronic states $g,e$ is calculated on the spatial grid extending from $R_{min}$ to $L_R$, as:
\begin{equation}
P_{g,e}(t) = \int_{R_{min}}^{L_R} | \Psi_{g,e}(R',t) |^{2} dR'
\label{eq:p1gt}
\end{equation}
The spatial grid is chosen such as at every instant $t$ the total population  is normalized at 1 on the grid ($P_{\Sigma}(t)+P_{1_g}(t)=1$). At t=0 the
population is entirely in the ground state ($P_{\Sigma}(0)=1$). 
\end{itemize}

\subsection{Time scales related to the laser coupling and vibrational movement}
\label{sec:timescale}

The time scales relevant for the dynamics are related to the laser coupling and to the vibrational movements in the electronic potentials.  The spontaneous emission from the excited state is neglected, as the time evolution of hundreds picoseconds studied here is short compared with the spontaneous emission time of about 30 ns. 

We begin by defining the characteristic times connected to the laser coupling.

 A {\it local time-dependent Rabi period} can be associated 
with the laser coupling $W_Lf(t)$ between the two electronic states \cite{chirpPRA04}:
\begin{eqnarray}
T_{Rabi}(R,t)=\frac{\hbar \pi}{\sqrt{(W_Lf(t))^2+\Delta^2(R)}},
\label{eq:Rabi}
\end{eqnarray}
where 2$\Delta(R)=|V_{1_g}(R)-V_{\Sigma}(R)|$ is the local detuning (see \fref{fig1_pot}). Such a characteristic time is relevant if the impulsive approximation remains valid on the whole duration of the pulse, i.e. if the relative motion of the two nuclei can be considered as frozen during the pulse duration.

For the rectangular pulse studied here the coupling remains constant in the time interval (100 ps, 300 ps), so we can refer in the analysis at a {\it local Rabi period} associated to the constant coupling $W_L$:
\begin{equation}
 T^L_{Rabi}(R)=\frac{\hbar \pi}{ \sqrt{W^2_L+\Delta^2(R)}}
\label{localRabi}
\end{equation}
This local Rabi period has its maximum at the potentials crossing ($T^L_{Rabi}(R_c)$=1.27 ps), diminishing with the increasing of the local detuning (for example $T^L_{Rabi}(89 \ a_0)$=0.24 ps). 

One can also associate a Rabi period with the {\it beating induced by the coupling $W_L$ between two specific vibrational states}, one belonging to the excited electronic state, and the other to the ground state. Indeed, in equation \eref{tschreq} the wavepackets  can be developed as superpositions of vibrational  wavefunctions $\{| \chi_{v_e,v_g}(R)>\}$ with eigenenergies $E_{v_e,v_g}$, corresponding to each electronic Hamiltonian ${\bf \hat H_{e,g}}={\bf \hat T}+{\bf \hat
  V_{e,g}}$  (${\bf \hat H_{e,g}}|\chi_{v_e,v_g}>=E_{v_e,v_g}|\chi_{v_e,v_g}>$):
\begin{eqnarray}
\Psi_{1_g}(R,t)= \sum_{v_e} c_{v_e}(t) \exp (-\frac{\rmi}{\hbar}E_{v_e}t)\chi_{v_e}(R), \label{superpos1} \\
 \Psi_{\Sigma}(R,t)= \sum_{v_g} c_{v_g}(t) \exp(-\frac{\rmi}{\hbar}E_{v_g}t)\chi_{v_g}(R),
\label{superpos2}
\end{eqnarray}
Supposing only two vibrational states, $c_{v_e}(t) \exp (-\frac{\rmi}{\hbar}E_{v_e}t)\chi_{v_e}(R)$ and $c_{v_g}(t) \exp(-\frac{\rmi}{\hbar}E_{v_g}t)$ $\chi_{v_g}(R)$, respectively, with $|c_{v_e}(t)|^2+|c_{v_g}(t)|^2=1$,
$|c_{v_g}(0)|^2=1$, $|c_{v_e}(0)|^2=0$, which are coupled by $W_L$ in equation \eref{tschreq}, one obtains for the oscillating population in the excited state:
\begin{eqnarray}
|c_{v_e}(t)|^2= \frac{|W_L<\chi_{v_e}| \chi_{v_g}>|^2}{(\hbar \Omega_{v_e,v_g})^2}
\sin^2(\Omega_{v_e,v_g} t),\\
\hbar \Omega_{v_e,v_g} = \sqrt{ |W_L<\chi_{v_e}| \chi_{v_g}>|^2 + [(E_{v_e}-E_{v_g})/2]^2 }.
\end{eqnarray}
The corresponding Rabi period will depend on the overlap $<\chi_{v_e}| \chi_{v_g}>$ of the vibrational functions:
\begin{equation}
T^L_{v_e,v_g}=\frac{\pi}{\Omega_{v_e,v_g}}
\label{vRabi}
\end{equation}

 The time scale associated with the {\it vibrational motion} of a vibrational level $v$ with binding energy $E_v$ in an electronic potential is:
\begin{equation}
T^{vib}_v=\frac{2 \pi \hbar}{|E_{v+1} - E_{v}|}.
\label{tvib}
\end{equation}
To estimate the coupling influence on the dynamics, the time scales \eref{localRabi} and \eref{vRabi} associated with the laser coupling shall be compared with the vibrational periods \eref{tvib} of levels in the ground and excited potentials. In the case treated here, the Rabi periods associated with the laser coupling are of the order of picosecond, being much smaller than the characteristic vibrational periods implied in the problem (tens or hundred ps), which indicates a case of {\it strong coupling} \cite{vatasescu01}. 

\begin{figure} 
 \center
 \includegraphics[width=0.7\columnwidth]{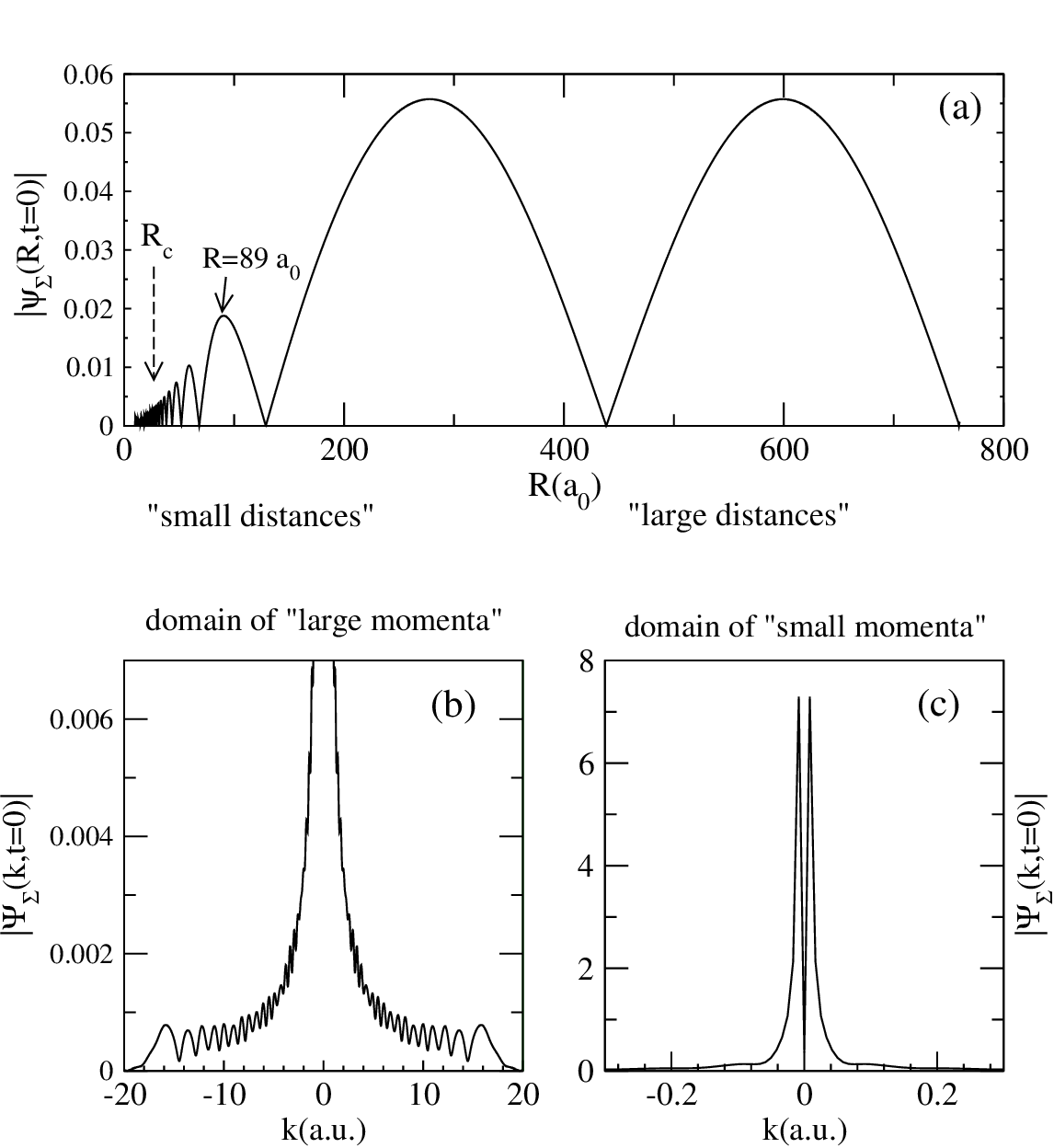}
          \caption{Initial state of the photoassociation process of energy $E_0$: stationary continuum state in the $a^3\Sigma_{u}^{+}$ potential, calculated and normalized to 1 in a box of length $L_R \approx 760$ a$_0$, corresponding to the temperature  $E_0/k_B$=0.11 mK. (a) Position representation $|\Psi_{\Sigma}(R,0)|$. (b) and (c) Momentum representation $|\Psi_{\Sigma}(k,0)|$ shown in the domains of ``large momenta`` ($|k| <$ 20 a.u.) and ``small momenta`` ($|k|<$ 0.2 a.u.), respectively.}
\label{fig3abc_psi0}
 \end{figure} 

\subsection{Initial state: spatial and momentum representations}
\label{sec:initial}

The representation of the initial state in a wavepacket treatment of the cold atoms photoassociation is discussed in \cite{elianeepjd04}. 
For a low temperature collision ($T=$0.11 mK) and an excitation process taking place at short distances ($R_c=29.3 \ a_0$), the initial state of the photoassociation process has to be represented using stationary collision states. We shall simulate the photoassociation dynamics using as initial state a continuum state of the ground electronic channel, having the energy $E_0=k_B T$. The results obtained can be used to estimate a photoassociation probability for an ensemble of cold atoms in thermal equilibrium at temperature $T$ (see \Sref{sec:endpulse}). 

In our numerical method, the initial continuum state is calculated as one of the eigenstates of the ground electronic state Hamiltonian, through the Sine Grid Representation \cite{willner04}, in a box of length $L_R$. The method introduces a discretization of the continuum which supply continuum states having a node at the boundary of the box (as the sine basis functions used in representation). Then it is possible to have a continuum delocalized state as initial state in the wavepackets simulation of the photoassociation dynamics. The MSG method allows the use of large spatial grids on which the wavepackets dynamics in the range of distances relevant for the problem can be followed for long propagation times without being influenced by the external boundary of the box \cite{elianeepjd04}. 

Here the initial state $\Psi_{\Sigma}(R,0)$ (see the (a) panel in \fref{fig3abc_psi0}) is chosen to be a continuum wavefunction of the
$a^3\Sigma_{u}^{+}$(6s,6s) potential, of energy $E_0=7.6908 \times 10^{-3}$
cm$^{-1}$ corresponding to a temperature $T=E_0/k_B=$0.11 mK. 
This wavefunction is calculated through the MSG method \cite{willner04,elianeepjd04} in the box of length $L_R \approx 760$ a$_0$, with a node at the end of the
grid, and normalized to 1 on the grid. To obtain the normalization per unit energy
 the populations have to be multiplied by the density of
states in the box at $E_0$, $[(\rmd E / \rmd n)|_{E_0}]^{-1}$ 
\cite{elianeepjd04}. The energy resolution for
neighbouring eigenstates in the box at $E_0$ is $\delta
E|_{E=E_0}= (\rmd E / \rmd n)|_{E_0} =0.632 \times 10^{-4}$
cm$^{-1}$, corresponding to $[\delta
E|_{E=E_0}]/k_B$=0.09 mK.

\begin{figure} 
\center
 \includegraphics[width=0.8\columnwidth]{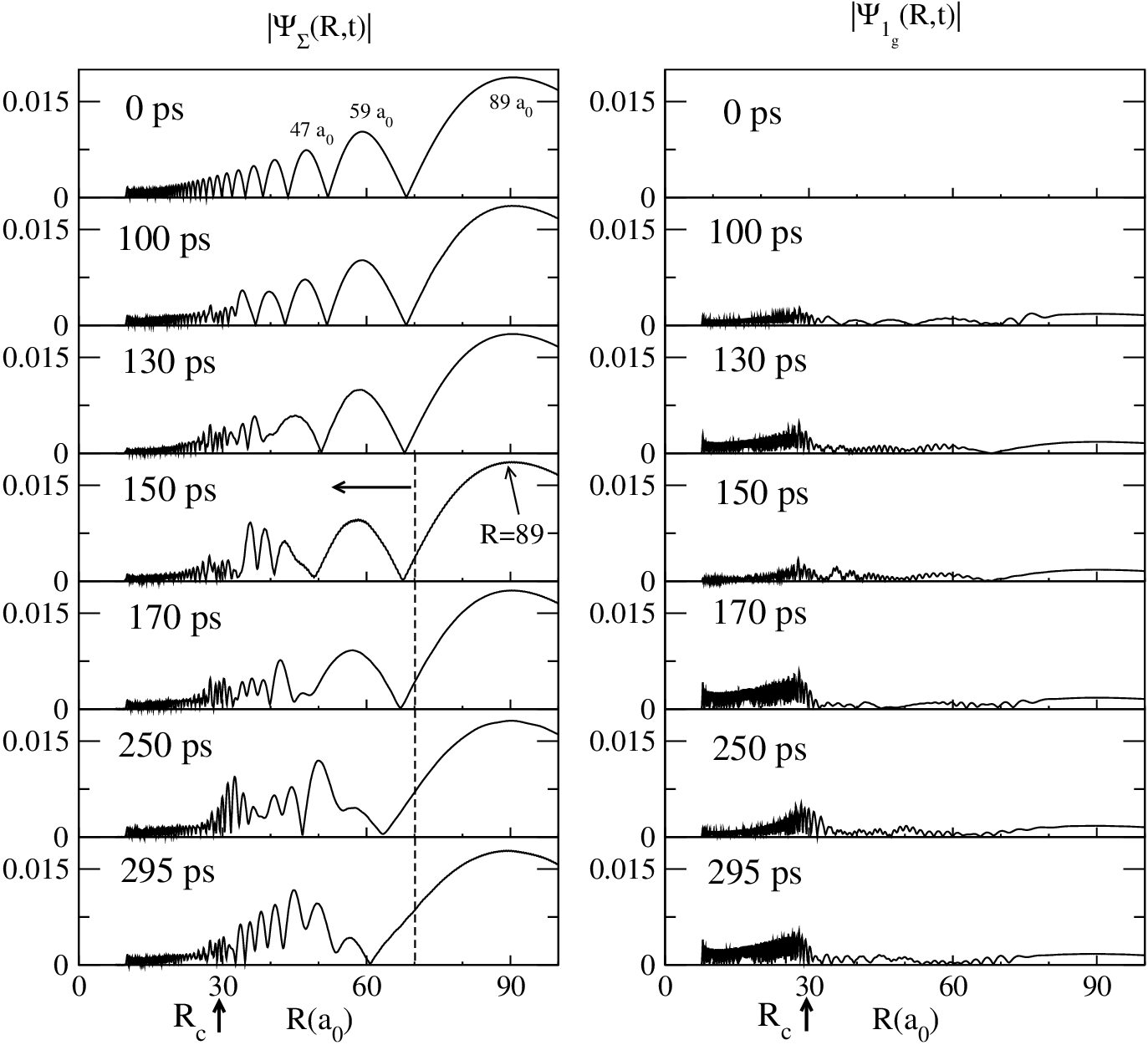}
  \caption{Time evolution of the wavepackets in the position representation: $|\Psi_{\Sigma}(R,t)|$ (left column) and $|\Psi_{1g}(R,t)|$ (right column). The vertical dashed line in the left column marks the part of the $a^3\Sigma_{u}^{+}$ wavepacket which is strongly accelerated inside the potential during the pulse.}
   \label{fig4_wfR}
\end{figure} 

Figures \ref{fig3abc_psi0}(b) and (c) show the amplitude of the initial wavefunction in the momentum representation, $|\Psi_{\Sigma}(k,0)|$. The wavefunction amplitude is mainly
localized in the domain of ``small momenta'', $|k|<$ 0.06 a.u.
(\fref{fig3abc_psi0}(c)), which corresponds to the large distances domain ($R>100$ a$_0$) in the position representation. A picture from the domain of ``large
momenta'' ($|k| <$ 20 a.u.), where the
wavefunction amplitude is much smaller, is displayed in \fref{fig3abc_psi0}(b).  This domain of momenta
corresponds to the domain of ``small distances'' in
\fref{fig3abc_psi0}(a). We are interested to observe the changes appearing in both domains during the time evolution. The
negative $k$ values mean momenta oriented to the inner wall of an electronic potential, and the positive $k$ values momenta oriented to large distances.

\begin{figure} 
\center
 \includegraphics[width=0.8\columnwidth]{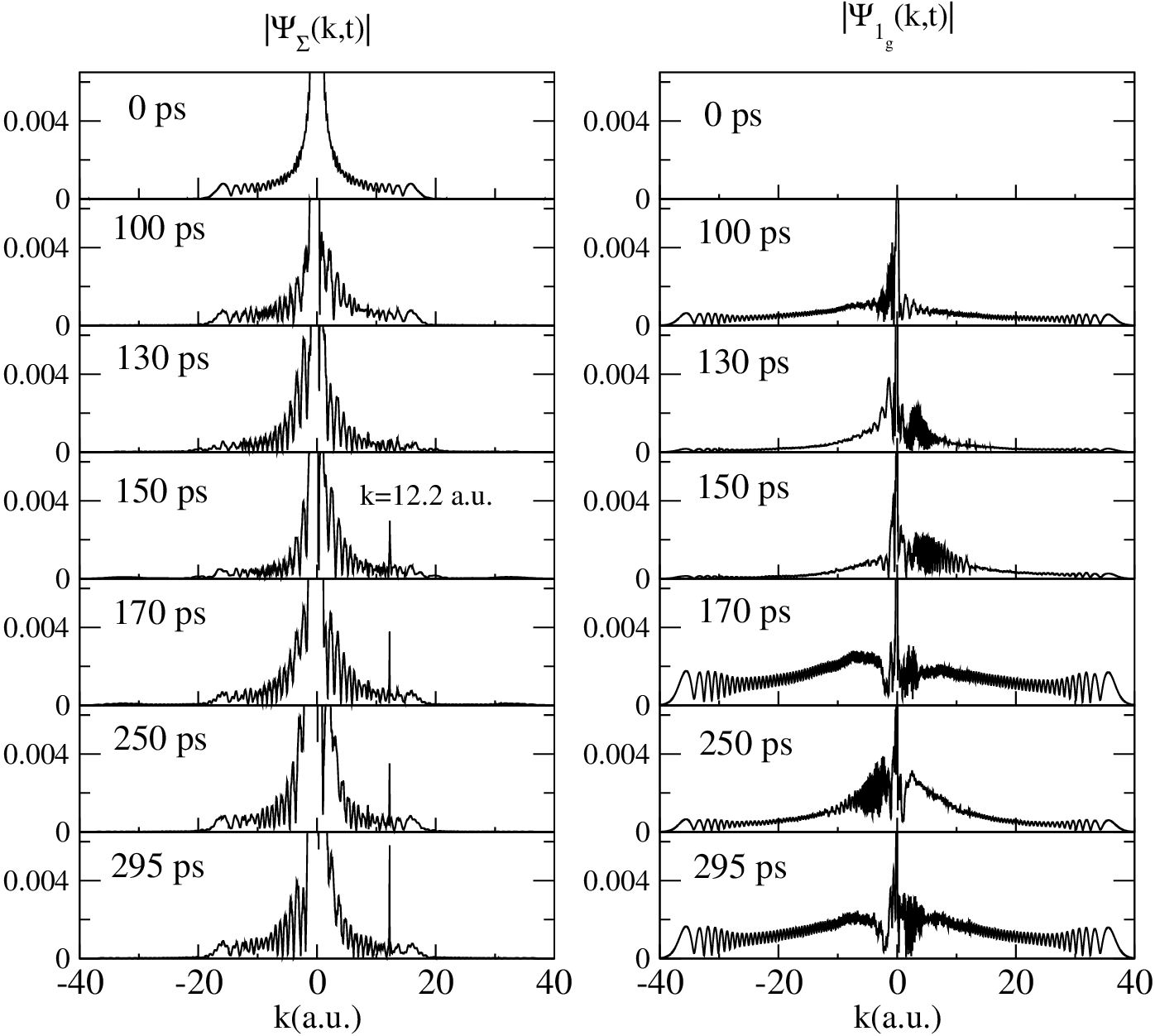}
          \caption{Time evolution of the wavepackets in the momentum representation $|\Psi_{\Sigma}(k,t)|$ (left column) and $|\Psi_{1g}(k,t)|$ (right column).}
          \label{fig5_wfP}
\end{figure} 

\subsection{Wavepackets evolution during the pulse: spatial and momentum representations}
\label{sec:timeev}

\begin{figure} 
\center
 \includegraphics[width=0.8\columnwidth]{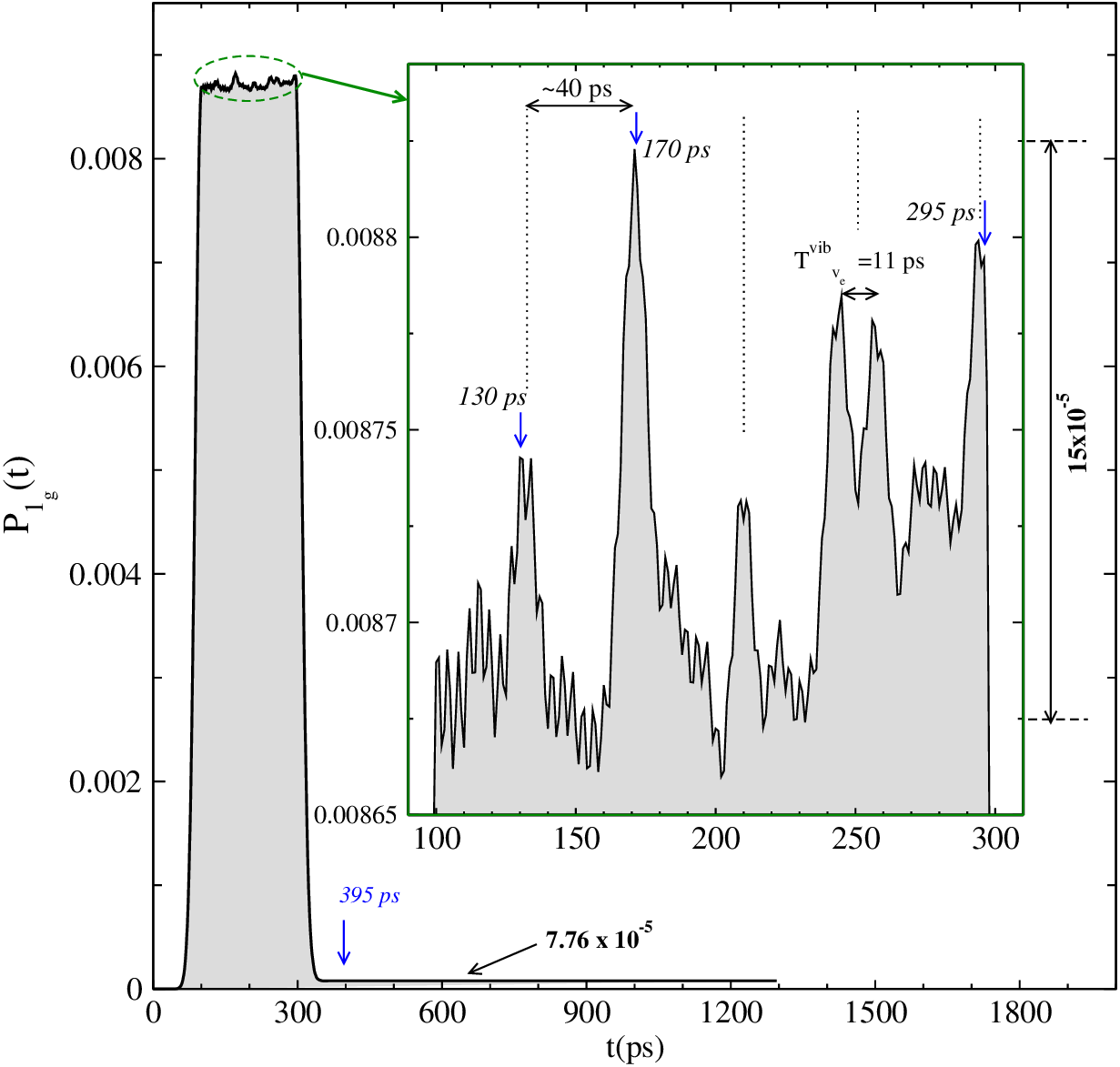}
          \caption{Time evolution of the population in the excited
          state, $P_{1g}(t)$, in the photoassociation with the pulse $W_L f(t)$ (whose envelope f(t) is represented in \fref{fig2_pulse}). The inset shows the
          details of the time evolution during the constant coupling
          $W_L$ (between 100 ps and 295 ps).}
          \label{fig6_norm1g}
\end{figure}

The evolution of the wavepackets during the pulse is illustrated in figures \ref{fig4_wfR} and \ref{fig5_wfP}, showing the $a^3\Sigma_{u}^{+}$ and $1_g$ wavepackets in the spatial and momentum representations, respectively. The dynamics will be analysed in order to understand the vibrational movements inside each channel and the exchange of population between the electronic channels (\fref{fig6_norm1g}). 

In the $1_g$  potential, the excited wavepacket extends on the whole spatial grid, showing that bound and continuum levels are populated
{\it during the pulse}. The vibrational movement of the population occupying bound states with outer turning points in the crossing region ($R_c = 29.3 \ a_0$) can be well observed in the right columns of figures \ref{fig4_wfR} and \ref{fig5_wfP}. The large amplitude of the wavepacket in the zone of big momenta (t=170, 250, 295 ps  in the right column in \fref{fig4_wfR}) is equivalent with a strong presence of population at $R<R_c$ at the same moments. {\it After the pulse}, only bound levels remain populated in $1_g$ (\fref{fig8_1g395} and \sref{sec:endpulse}).

In the $a^3\Sigma_{u}^{+}$ potential, the wavepacket moves progressively to the inner region (left column of \fref{fig4_wfR}). The vertical line in the same figure marks the separation in two spatial domains, which are differently affected by the pulse: at small and intermediate distances $R < 70 \ a_0$, the wavepacket is accelerated and deformed, but at large interatomic separations $R > 80 \ a_0$ the action of the pulse can be considered as impulsive.

The wavepackets dynamics in the momentum representation (\fref{fig5_wfP}) makes visible an unexpected feature which appears from t=150 ps in the time evolution of the ground state wavepacket $\Psi_{\Sigma}(k,t)$ at $k \approx 12.2$ a.u., and whose intensity increases until the end of the pulse. This result will be analysed in \sref{sec:kfeatures}.

\section{Analysis of the population transfer during the pulse}
\label{sec:poptransfer}
We shall analyse the time evolution of the population transferred by the pulse in the excited  state, $P_{1_g}(t)$, calculated with formula \eref{eq:p1gt} and displayed in \fref{fig6_norm1g} (the population in the ground state has a complementary evolution, as $P_{\Sigma}(t)+P_{1_g}(t)=1$, with $P_{\Sigma}(0)=1$). The figure shows that from the large amount of population $P_{1_g}(t)$ transferred  during the pulse, which is of the order of $8.7 \times 10^{-3}$, only $7.77 \times 10^{-5}$ remains at the end of the pulse. As it was discussed in the previous section, the large population transfer in the excited state during the pulse is related to the occupation of continuum states at large distances, in levels which do not rest populated after the pulse \cite{chirpPRA04,elianeepjd04}. On the contrary, what it is interesting for us is the population in bound vibrational levels, counting as cold molecules formation.

\begin{table}
\caption{\label{tab:chartimes} Vibrational levels $v_e$ of the excited state (with energies
  $E_{v_e}$ and vibrational periods $T^{vib}_{v_e}$) and  $v_g$ of the
  ground state (with energies $E_{v_g}$ and vibrational periods $T^{vib}_{v_g}$), populated during the pulse and giving the
  quantum beats shown in the inset of \fref{fig6_norm1g}. $\{v_e\}$ are 
  $1_g$ vibrational levels located around the crossing of the 
  diabatic potentials and $\{v_g\}$ are
  vibrational levels in the $a^3\Sigma_{u}^{+}$ potential giving the
  biggest overlaps $<\chi_{v_g} |\chi_{v_e}>$ with every $v_e$
  level. The last two columns show the energy differences $|E_{v_g}-
  E_{v_e}|$ and the characteristic times $T^L_{v_g,v_e}$,
  calculated with formula \eref{vRabi}. The energy origin
  is the dissociation limit $E_{6s+6p_{3/2}}$ of the
$1_g(6s,6p_{3/2})$ potential.} 
\begin{indented}
\lineup
\item[]\begin{tabular}{@{}*{9}{l}}
\br
$v_e$&$T^{vib}_{v_e}$&$E_{v_e}$&$v_g$&$T^{vib}_{v_g}$&$E_{v_g}$&$<\chi_{v_g} |\chi_{v_e}>$&$|E_{v_g}- E_{v_e}|$&$T^L_{v_e,v_g}$\cr
&(ps)&(cm$^{-1}$)&&(ps)&(cm$^{-1}$)&&(cm$^{-1}$)&(ps)\cr
\mr
141&10.6&-144.05&44&40&-143.27&\0\0\00.10&0.78&12.7\cr
&&&45&49&-142.43&\0\0\00.22&1.62&5.5\cr
&&&46&62&-141.75&\0\0\00.16&2.30&7\cr
&&&47&80&-141.21&\0\0\00.13&2.84&7.5\cr
\mr
142&10.8&-140.90&44&40&-143.27&\0\0\00.08&2.37&10.5\cr
&&&45&49&-142.43&\0\0\00.20&1.53&6\cr
&&&46&62&-141.75&\0\0\00.19&0.85&6.6\cr
&&&49&155&-140.48&\0\0\00.10&0.42&12.5\cr
\mr
143&11&-137.81&44&40&-143.27&\0\0\00.06&5.46&9\cr
&&&45&49&-142.43&\0\0\00.19&4.62&4.5\cr
&&&46&62&-141.75&\0\0\00.22&3.94&4.8\cr
&&&49&155&-140.48&\0\0\00.11&2.67&8.5\cr
\mr
144&11.1&-134.77&44&40&-143.27&\0\0\00.05&8.50&3.9\cr
&&&45&49&-142.43&\0\0\00.17&7.66&3.8\cr
&&&46&62&-141.75&\0\0\00.23&6.98&2.5\cr
&&&48&108&-140.79&\0\0\00.12&6.02&5\cr
\br
\end{tabular}
\end{indented}
\end{table}

The vibrational levels of $1_g$ and $a^3\Sigma_{u}^{+}$ predominantly populated by the pulse were identified by calculating the probabilities  $P_{v_e,v_g}(t)=|<\Psi_{e,g}(R,t)|\chi_{v_e,v_g}(R)>|^2$ that a certain vibrational level of the excited or ground electronic state ($v_e$ or $v_g$) to be populated at an instant $t$. 
Two kinds of bound vibrational levels are mainly populated during the
pulse in the $1_g$ state:
\begin{itemize}
\item levels around the crossing of the diabatic potentials:
$v_e$=141-144, with vibrational energies $E_{v_e}=$-144.5,
-141.39, -138.29, -135.24  cm$^{-1}$, lying in a domain of about 10 cm$^{-1}$ containing the
crossing energy $V_c=-143$ cm$^{-1}$, and with outer turning points
of the vibrational wavefunctions lying around $R_c = 29.3$
a$_0$. These levels have vibrational periods of about 11 ps (see \tref{tab:chartimes}).
\item levels with $v_e \geq
  241$ ($E_{v_e=241} = -8.818$ cm$^{-1}$), having vibrational wavefunctions lying at much larger distances ($R > 60$ a$_0$). From 
  these levels populated at large distances, 
  only the levels $v_e=$244 up to 248 remain notably populated after
  the pulse. Their vibrational periods are in the range 116 up to 130 ps.
\end{itemize}

The inset of \fref{fig6_norm1g} shows the evolution of  $P_{1_g}(t)$ in the time interval of constant coupling $W_L$ (between 100 ps and 295 ps).
The oscillating features of $P_{1_g}(t)$ during this time interval come from the exchange of population between vibrational levels of $1_g$ and $a^3\Sigma_{u}^{+}$  located around the crossing ($v_e$=141 up to 144 in the excited state, and $v_g$=44 up to 49 in the ground state, see \tref{tab:chartimes}), without major contribution from the population transferred at large
distances. Indeed, as it is marked on
the inset of \fref{fig6_norm1g}, the energy domain covered by these oscillations is about $15 \times10^{-5}$, comparable with the final population $P_{1_g}(t=395
ps)=7.77 \times 10^{-5}$ in bound levels. Also, the variations of the total population $P_{1_g}(t)$ between instants as t=130, 150, 170 ps, etc.  are very closed to the variations of the population  in levels around the
 crossing, given by the sum $P^{141-144}_{1_g}$(t)=$\sum_{v_e=141}^{144}P_{v_e}(t)$.
  Comparatively, the probability for the population of the levels $v_e$=244-248,
$P^{244-248}_{1_g}$(t)=$\sum_{v_e=244}^{248}P_{v_e}(t)$, shows much smaller variations. 

The quantum beats appearing in \fref{fig6_norm1g} are related to the characteristic times of the dynamics: the vibrational movement in each potential well and the beating between the two coupled wavepackets, which are superpositions
of vibrational functions corresponding to each electronic potential, as in \eref{superpos1} and \eref{superpos2}. \Tref{tab:chartimes} gives a list of energies $E_v$, vibrational periods $T^{vib}$, and characteristic beating 
times $T^L_{v_g,v_e}$ between levels in the ground and excited electronic states which contribute significantly in the exchange of populations between the two coupled channels; it also contains the overlaps $<\chi_{v_g} |\chi_{v_e}>$  with  $a^3\Sigma_{u}^{+}$ levels. Only the $v_g$ levels having the biggest overlaps with a given $v_e$ level are shown: these are the levels $v_g$=44 up to 49, whose
vibrational wavefunctions have the outer turning points at distances
$R<35 \ a_0$ in the $a^3\Sigma_{u}^{+}$ potential, and which are strongly
populated by the pulse. 

The population transfer between the two electronic channels is regulated by two time scales: a longer one, related to the vibrational movements inside the potential wells, and a shorter one, related to the Rabi coupling.
The longer scale reflects the influence of the vibrational movements on the exchange of population between the two channels: the exchange is maximal when the amplitudes of the two wavepackets have a significant overlap, which, in the case of wavepackets vibrating in two different potential wells, arrives when both wavepackets have important localization probabilities in the crossing region \cite{vatasescu01}. When one of the wavepackets vibrates inside its potential, the transfer is generally diminished. The levels in the ground state have vibrational periods ($>$ 40 ps) much longer than the levels in the excited state (about 11 ps). This explains the period of 40 ps appearing in the $P_{1_g}(t)$ oscillations, which  coincides with the vibrational period of the level $v_g$=44 in the $a^3\Sigma_{u}^{+}$ ground state,  whose energy -143.27 cm$^{-1}$ is very close to the crossing energy $V_c$. On a much shorter scale, the transfer is guided by the strong laser coupling, which couples differently the implied levels. Table \ref{tab:chartimes} shows  characteristic 
times $T^L_{v_g,v_e}$ of beating (calculated with formula \ref{vRabi}) varying from 2.5 to 12.7 ps for coupled pairs of vibrational levels in the ground and excited states. Comparing these times with the vibrational periods of the concerned levels, the strength of the coupling appears as varying very much among pairs of levels. The times scales given by $T^L_{v_g,v_e}$ are indicative for the short Rabi times appearing during the dynamics, for example the periods of 3 up to 4 ps of the fast oscillations in $P_{1_g}(t)$.

During the pulse the population accumulates in the $1_g$ state, such as in \fref{fig6_norm1g}  
appear not only the population beatings between the two channels, but also traces of the vibrational dynamics inside the excited potential: around t=250 ps, the two peaks of $P_{1_g}(t)$ are separated by a time interval of
$\approx$ 11 ps, which is the vibrational period of the $v_e=142,143$
levels located around the crossing: $T^{vib}_{v_e=142,143} \approx 11$  ps.

\subsection{The light-induced (adiabatic) potentials} 
\label{sec:lightpot}
\begin{figure} 
\center
 \includegraphics[width=0.7\columnwidth]{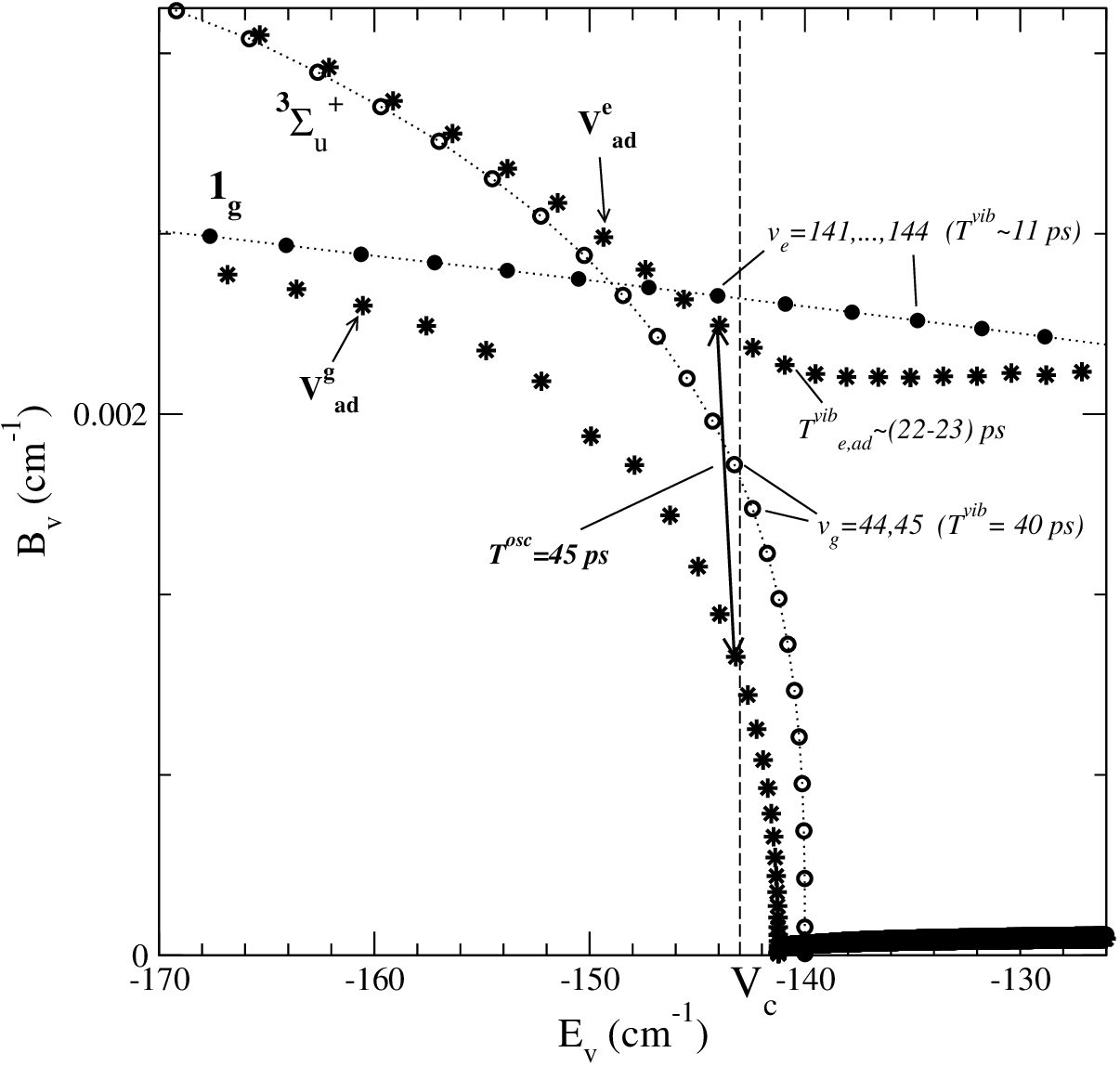}
          \caption{Rotational constants $B_{v}=\langle \chi_{v}
|\hbar^2/(2 \mu  R^2)| \chi_{v}\rangle$ as functions of the  vibrational
energies $E_v$, for vibrational levels of the $1_g$
 and $a^3\Sigma_{u}^{+}$ diabatic potentials (full and empty circles,
 respectively), and of the adiabatic potentials $V_{ad}^g$, $V_{ad}^e$ (stars). The vertical line at $V_c=-143$ cm$^{-1}$ indicates the crossing region.}
          \label{fig7_bvc}
 \end{figure}
The mechanism of the population transfer during the pulse is enlightened if one analyses the light-induced (adiabatic) potentials.
In \fref{fig1_pot} are represented both the diabatic
potentials  $V_{1_g,\Sigma}(R)$ and the adiabatic ones, $V_{ad}^e$ and
$V_{ad}^g$, obtained from the diagonalization of the
2x2 potential matrix ${\bf \hat V_{el}}$ with constant coupling ${\bf
 \hat W_L}$ as non-diagonal term:
\begin{eqnarray}
({\bf \hat V_{el}}+{\bf \hat W_L} )|_{diab}=\left(\begin{array}{lc}
 V_{1_g}(R)  &
W_L  \\
 W_L  &
  V_{\Sigma}(R)
 \end{array} \right)
\label{diabmatrix}
\end{eqnarray}
\begin{eqnarray}
 ({\bf \hat V_{el}}+{\bf \hat W_L} )|_{adiab}=\left(\begin{array}{lc}
 V_{ad}^e(R)  &
0  \\
 0  &
  V_{ad}^g(R)
 \end{array} \right)
\label{adiabmatrix}
\end{eqnarray}
Expressions \eref{diabmatrix} and \eref{adiabmatrix} illustrate the diabatic and
adiabatic representations, respectively, of the ${\bf \hat
  V_{el}}+{\bf \hat W_L}$ operator. The coupling produces a strong deformation of the diabatic potentials $V_{1_g},
V_{\Sigma}$ around the crossing (\fref{fig1_pot}), but its influence goes far beyond the crossing region, and this can be
clearly seen if one compares the vibrational levels of the
diabatic potentials $V_{1_g},
V_{\Sigma}$  with those
of the  adiabatic ones $V_{ad}^e$, $V_{ad}^g$. Such a
comparison can be made using the  series of rotational constants $B_{v}=\langle \chi_{v}
|\hbar^2/(2 \mu  R^2)| \chi_{v}\rangle$ characterizing every
potential. The energies $E_{v}$ and rotational constants $B_{v}$ were computed by solving
numerically  the corresponding stationnary
Schr\"odinger equation $({ \bf \hat T}+{\bf \hat V}) |\chi_{v}>=E_v |\chi_{v}>$ using the
Mapped Fourier Grid Hamiltonian (MFGH) method \cite{slava99}. In \fref{fig7_bvc} are represented the rotational constants $B_v$ as functions of the
vibrational energies  $E_v$ for $a^3 \Sigma_u^+$ and $1_g$
diabatic potentials, as well as for the adiabatic
potentials $V_{ad}^e$, $V_{ad}^g$. We use the $B_v(E_v)$ functions to observe how the energies of the vibrational levels  in these potentials are distributed in a
domain lying between -170 and -128 cm$^{-1}$ around the crossing energy $V_c=$-143 cm$^{-1}$. These results show that the influence of the coupling is strongly felt in a large energy domain of several tens of cm$^{-1}$ around the crossing.

Also marked in \fref{fig7_bvc} are the vibrational levels populated during the pulse and relevant to the dynamics, together with their vibrational periods. We shall focus on the levels located around the crossing region. The levels $v_g=44, 45$ in the $a^3 \Sigma_u^+$ potential have vibrational periods $T^{vib}_{v_g}=40, 49$ ps. In the same energy region, the levels of the $V_{ad}^g$ potential have bigger vibrational periods $T^{vib}_{g,ad}=43,58$ ps. Also,
if the vibrational period of the $1_g$ levels located around the crossing, $v_e=141$ up to $144$, is $T^{vib}_{v_e} \approx 11$ ps, in the adiabatic potential $V_{ad}^e$ the levels belonging to the same energy domain have a vibrational period twice bigger: $T^{vib}_{e,ad} \approx 23$ ps. Then, as it could be expected from the shape of the $V_{ad}^e$ and $V_{ad}^g$ (\fref{fig1_pot}), the vibration in the adiabatic potentials is slowed down in the crossing region, influencing the population transfer between the two channels. 
The slowing down of the vibrational movement in the excited state lead to longer periods for the population exchange between the two channels and constitutes
a mechanism for increasing progressively the population in the excited state. 

The analysis of the adiabatic potentials can be used to extract characteristic times for the population  exchange between the two channels. In an analogy with a two coupled states system, the Rabi beatings in the exchange of population is related to the Bohr frequency of the coupled system, which here can be found from the typical frequencies $|E_{ad}^e-E_{ad}^g|/2h$, connected with the new energies $E_{ad}^e$, $E_{ad}^g$ of the levels in the coupled system. In \fref{fig7_bvc} we show that for levels located in the crossing region such a characteristic time
$T_{osc}=2 \pi \hbar / |E_{ad}^e-E_{ad}^g|$ is about 45 ps, close to the period of $\approx 40$ ps of the large oscillations in \fref{fig6_norm1g}.

\section{Results at the end of the pulse}
\label{sec:endpulse}
\begin{figure}
\center
\includegraphics[width=0.6\columnwidth]{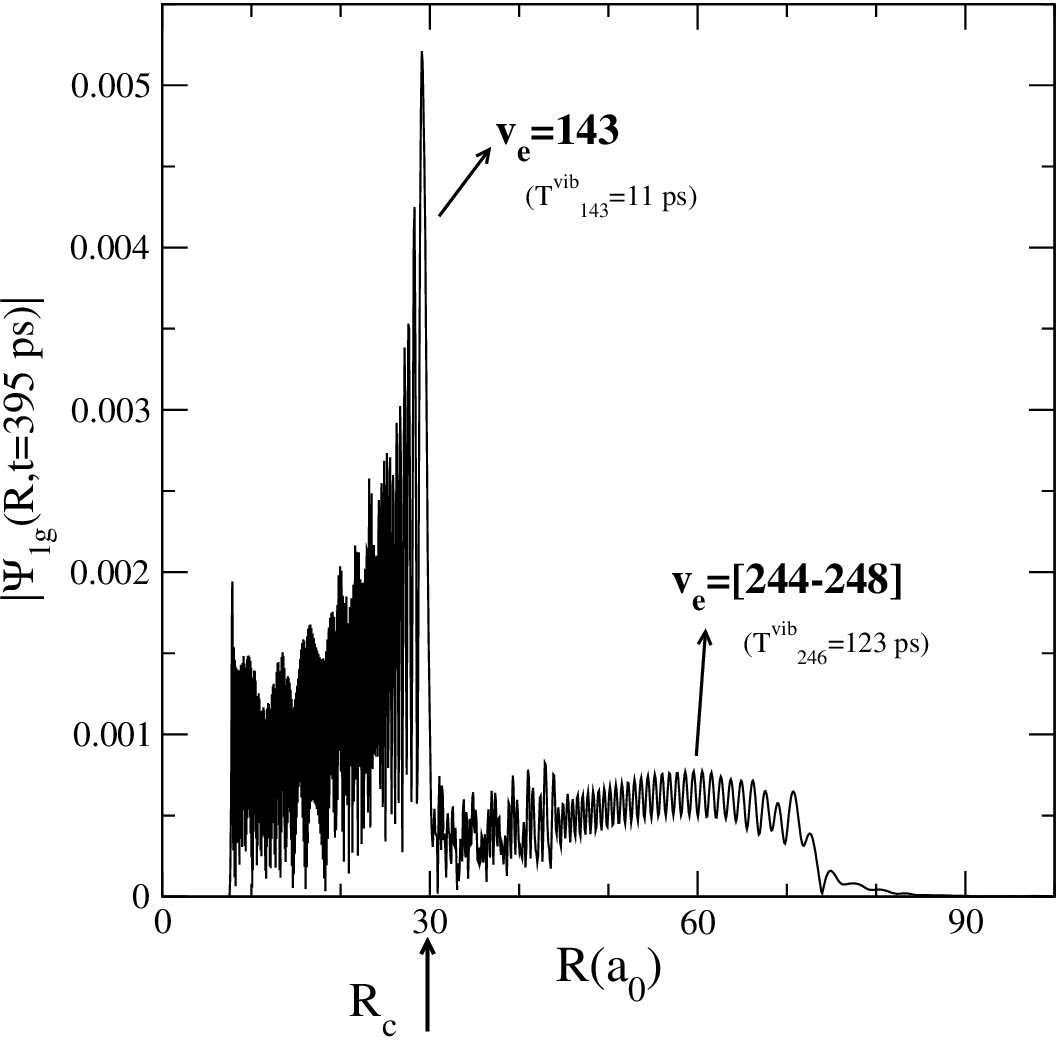} 
\caption{$1_g$  wavefunction at the end of the pulse.}
\label{fig8_1g395}
\end{figure}

Figures \ref{fig8_1g395} and  \ref{fig9abc_psiS395} show the wavepackets $1_g$ (in position representation) and $a^3\Sigma_{u}^{+}$ (in position and momentum representations) at the end of the pulse (t=395 ps). The main characteristics of the results will be discussed in the following. 

\subsection{Formation of strongly bound cold molecules in $a^3\Sigma_{u}^{+}$ and  $1_g$ electronic states}
\label{sec:finpop}
The strong coupling between the ground and excited states creates an interesting result at the end of the pulse: the population in bound levels of the
  ground state (mainly  six levels, $v_g=$47 up to 52),  $P^{47-52}_{\Sigma}$(t=395 ps)=2.83 $\times$
10$^{-4}$, is much bigger than the population photoassociated in the excited state, P$_{1_g}$(t=395 ps)=0.78 $\times$ 10$^{-4}$. Some of the vibrational levels populated in $a^3\Sigma_{u}^{+}$, for example $v_g=$47,48,49, have  wavefunctions
localized at distances $R < 35$ a$_0$, being then strongly bounded. 

\begin{figure}
\center
 \includegraphics[width=0.8\columnwidth]{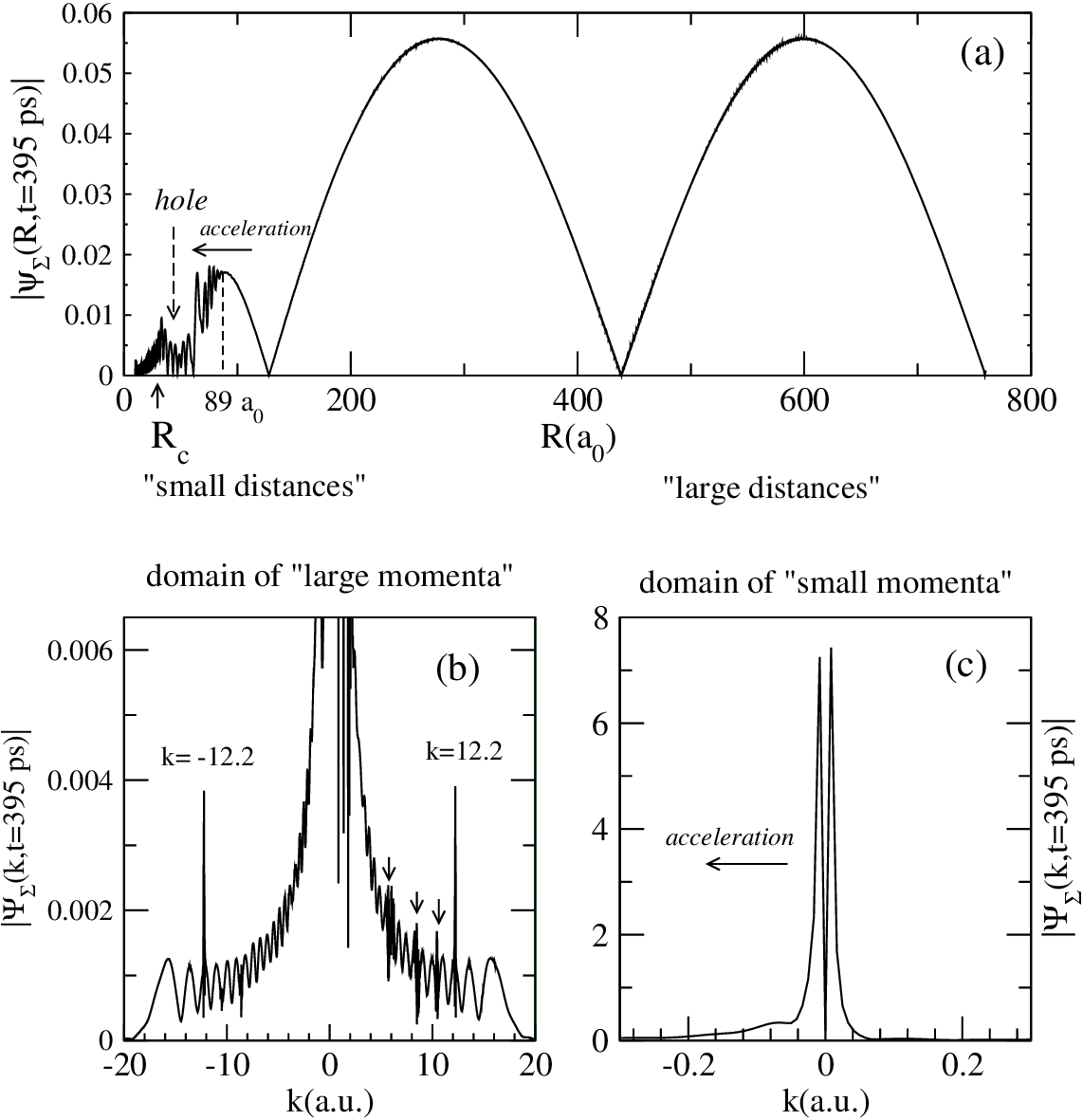}
\caption{$a^3\Sigma_{u}^{+}$  wavepacket at the end of the pulse, t=395 ps.  (a) Position representation $|\Psi_{\Sigma}(R,t=395 ps)|$. (b) and (c) Momentum representation $|\Psi_{\Sigma}(k,t=395 ps)|$ in the domain of ``large momenta`` ($|k| <$ 20 a.u.) and  ``small momenta`` ($|k| <$ 0.2 a.u.), respectively.}
\label{fig9abc_psiS395}
\end{figure}

\begin{figure}
\center
 \includegraphics[width=0.7\columnwidth]{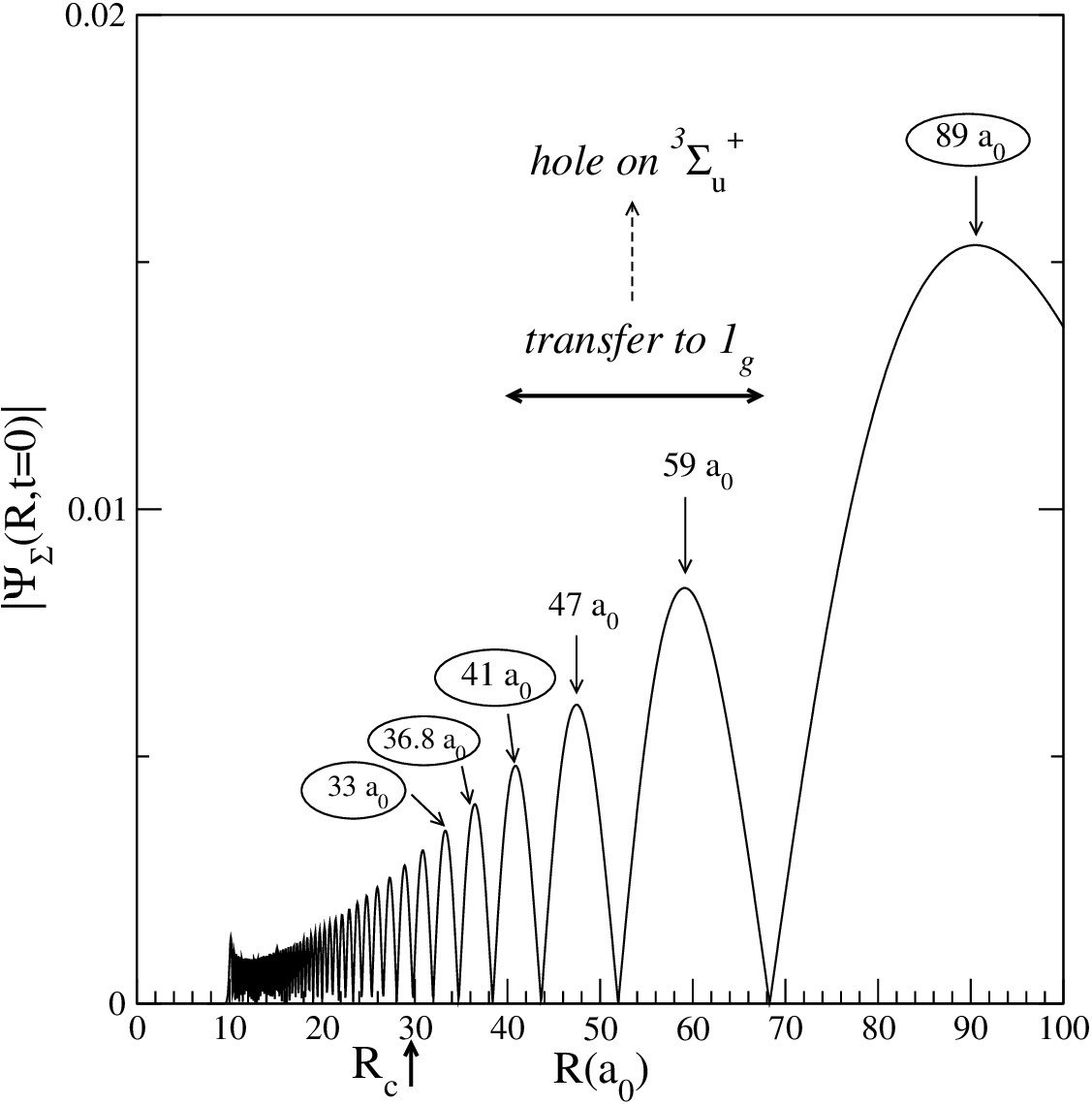}
\caption{Details of the initial wavefunction on the ground state, $|\Psi_{\Sigma}(R,t=0)|$. The arrows show local maxima of the wavefunction. The encircled ones give contributions to the off-resonance cycling of population between the two channels, bringing momentum to the final ground state wavepacket. On the contrary, ground state population from the local maxima at R=47 a$_0$ and 59 a$_0$ is tranferred to the excited state 1$_g$ (\fref{fig8_1g395}), creating a hole in the final $a^3\Sigma_{u}^{+}$ wavepacket (\fref{fig9abc_psiS395}a).}
\label{fig10_psir0max}
\end{figure}

The final population in the $1_g$ excited state (\fref{fig8_1g395}) is entirely in bound vibrational states. A superposition of two kinds of vibrational levels is created, showing two mechanisms in the population transfer: {\it at resonance}, where
mainly one level rests populated, $v_e$=143 (with outer turning point at $\approx 30$ a$_0$), whose population represents 82$\%$ P$_{1_g}$(t=395 ps), and {\it off-resonance} where several
vibrational levels with outer turning points around $R \approx 70$
a$_0$ ($v_e$=244 up to 248, representing $15\%$
P$_{1_g}$(t=395 ps)) are populated due to the strong coupling catching the
large amplitude of the initial wavefunction between 45 and 65 a$_0$
(\fref{fig10_psir0max}). This last kind of transfer creates a {\it hole in the ground state} in this domain of interatomic distances, as it is indicated in figures \ref{fig9abc_psiS395}(a) and \ref{fig10_psir0max}. Such a result reflects the specificity of the present photoassociation conditions of strong field and large detuning.

The results obtained for a total  population normalized at 1 on the grid allow
an estimation of the averaged probability
corresponding to a thermal distribution \cite{elianeepjd04} at the temperature
$T=E_0/k_B$=0.11 mK, as:
\begin{equation}
P{_{1_g}}(T) \approx P_{1_g}(E_0) (\frac{dE}{dn}|_{E_0})^{-1} \frac{k_B T}{Z}
\end{equation}
where $P_{1_g}(E_0)$=P$_{1_g}$(t=395 ps)=0.78 $\times$ 10$^{-4}$ is the $1_g$ probability obtained in the present calculation with an initial state of energy $E_0$,
$[(\rmd E / \rmd n)|_{E_0}]^{-1}$ is the density of states in the box of length $L_R$ at
$E_0$, $k_B$ is the Boltzmann constant, and $Z$ is the partition
function for a gas composed of non-interacting pairs of atoms in a
volume $V$ (with $\mu$ the reduced mass of the diatom): $Z=(2\pi
  \mu k_B T)^{3/2} V /h^3$.
We then obtain $ZP{_{1_g}}(0.11\  mK)=0.95 \times 10^{-4}$ and 
$ZP^{v_g=47-52}_{\Sigma}(0.11 \ mK)=3.44 \times 10^{-4}$.
For a number of $N$ atoms in a volume $V$, and taking into account the
spin degeneracy of the $Cs(6^2S)$ atomic state and of the initial
electronic state, the total number of molecules photoassociated in
$1_g$ per  pump pulse is \cite{elianeepjd04}:
$N_{mol,1_g}=\frac{N^2}{2}P{_{1_g}}(T)\frac{3}{4}$.
For a volume $V$=10$^{-3}$ cm$^3$ and a density of atoms
$N / V$=10$^{11}$ atoms/cm$^3$, the number of molecules obtained
at the end of the pulse are: $N_{mol,1_g}=0.3 \times 10^{-2}$ and
 $N_{mol,\Sigma,v_g=47-52}=1.1 \times 10^{-2}$. 

\subsection{Acceleration of the $a^3\Sigma_{u}^{+}$ wavepacket to small interatomic distances} 
\label{sec:accground}
During the pulse, the $a^3\Sigma_{u}^{+}$ slow packet is accelerated towards small interatomic distances R. The left column of \fref{fig4_wfR} shows the time evolution of the  $a^3\Sigma_{u}^{+}$ wavepacket, which advances progressively to the inner zone. Especially the part of the wavepacket occupying distances R $<$ 70 a$_0$  ¨feels¨ the acceleration to the crossing zone, where the diabatic potential $a^3\Sigma_{u}^{+}$ in $R^{-6}$ becomes the adiabatic one $V_{ad}^g$ decreasing in $R^{-3}$ (see \fref{fig1_pot}). On the contrary, the maximum of the wavefunction located at R $\approx$ 89 a$_0$ does not move during the pulse, but begin to be accelerated after the pulse. Similar observations can be made on \fref{fig9abc_psiS395} (a), showing  the $a^3\Sigma_{u}^{+}$ R-wavepacket at the end of the pulse (t=395 ps): the changes in the wavepacket amplitude are at distances R $<$ 89 a$_0$. The momentum representation of the $a^3\Sigma_{u}^{+}$ wavepacket at the same instant t=395 ps, in \fref{fig9abc_psiS395} (c), shows that, compared with the initial symmetric distribution (\fref{fig3abc_psi0} c),  the part corresponding to k$<$0 has moved to bigger $|k|$ values, which also indicates the gain of kinetic energy in the electronic potential in the movement to small distances. 

The creation of a ``hole`` in the ground state wavepacket at the end of the pulse (discussed previously and shown in \fref{fig9abc_psiS395}a)) is also a factor leading to a compression of population at short range, which after the pulse acts to increase the density of atom pairs at  short distances \cite{elihole07}. 

\subsection{Kinetic energy ``gains'' in the final $a^3\Sigma_{u}^{+}$ wavepacket as signatures of the maxima in the initial wavefunction continuum}
\label{sec:kfeatures}

In this section we shall analyse some peculiar features appearing in the ground state wavepacket $|\Psi_{\Sigma}(k,t)|$ during the time evolution and at the end of the pulse (\fref{fig9abc_psiS395}b)). The time evolution in the momentum representation (left column of \fref{fig5_wfP}), shows that, from t=150 ps, a line strikingly appears in the wavepacket amplitude $|\Psi_{\Sigma}(k,t)|$, at  k $\approx 12.2$ a.u. The kinetic energy associated with this momentum is  $\hbar^2 k^2 /2 \mu$$=134.8$ cm$^{-1}$$=|V_{1_g}(R)-V_{\Sigma}(R)|_{R=89 a_0}$, corresponding to the local difference in energy between $V_{1_g}(R)$ and
$V_{\Sigma}(R)$ potentials at $R=89 \ a_0$. The amplitude of the initial wavefunction $|\Psi_{\Sigma}(R,0)|$ has a maximum at this distance (\fref{fig3abc_psi0} a), which does not move during the time evolution, but on which small oscillations begin to be superposed (see the left column of \fref{fig4_wfR}). The period of these oscillations is $T_R=0.5$ a$_0$ $=2\pi/k$, corresponding to a plane wave $\exp (\rmi k R)$ with $k = 12.2$ a.u. The momentum width  of this $k$-feature in the  $|\Psi_{\Sigma}(k,t)|$ wavepacket is $\Delta k \approx 0.15$ a.u., which is consistent with the spread in distance $\Delta R \ge 3.3 \ a_0$  on $|\Psi_{\Sigma}(R,t)|$ around R $\approx 89 \ a_0$. We interpret this feature as corresponding to the kinetic energy of the population 
cycled back in the ground state from the excited state in the fast exchange of populations taking place around R $\approx 89\ a_0$ due to the strong coupling between the two electronic channels, with
$T^L_{Rabi}(89 \ a_0)$=0.24 ps.
At the end of the pulse, the line with k $\approx 12.2$ a.u. is accompanied in the $|\Psi_{\Sigma}(k,t=395 \ ps)|$  wavepacket by its negative value $k=-12.2$ a.u. (see \fref{fig9abc_psiS395}b), which has to be a technical artefact coming from the propagation of this large positive
  momentum component to large interatomic R distances, followed by reflection at the end of the grid.

  In fact, as it is shown in the Appendix, it seems that the momentum $k$ first appears associated with an ingoing plane wave travelling to small distances $(k<0)$, but it is fast reflected by the inner wall of the potential and transformed in a positive $k$ travelling to large distances, which is easy to be observed in the wavepackets evolution.  We have simulated the dynamics in the same conditions, but taking as initial wavepacket in the ground state a gaussian centered at R $= 89\ a_0$, and indeed we observed a peak at $k=-12.2$ a.u. appearing early in the time evolution.

$k=12.2$ a.u. is not the only value of momentum for which a feature appears in the $|\Psi_{\Sigma}(k,t)|$ wavepacket.  At various instants of the time evolution, other lines can
be observed, at smaller values of $k$. As these lines are embedded in the wavepacket, it becomes easier to distinguish  them for larger momenta where the wavefunction amplitude is smaller. In \fref{fig9abc_psiS395}b,
showing the $a^3\Sigma_{u}^{+}$ wavepacket amplitude in the domain of large momenta,
at the end of the pulse, other lines can be observed at the
values: k=12.2, 10.4, 8.6 and 6 a.u. They correspond to kinetic
energies $\hbar^2 k^2 / 2 \mu$=134.8, 98, 68.6 and 32.6
cm$^{-1}$. If one associates these kinetic energies to local
differences $\Delta V (R)$ between the electronic potentials,  then, according to the reasoning just exposed 
they correspond to local transitions from the excited to the ground state taking place at distances R $\approx$ 89, 43.9, 36.8 and 32 a$_0$, where maxima
of the initial wavefunction are located, as it appears in \fref{fig10_psir0max}. Other local maxima of the initial wavefunction, at R=47 a$_0$ and 59 a$_0$, do not
have a correspondent k-value in the final momentum distribution of
the  ground state wavepacket, but in this domain of
distances the population is not cycled back to the ground state, remaining transferred to the levels $v_e$=244 up to 248 of the excited state (see  \fref{fig8_1g395}). As a consequence, as we showed already, a hole is created in the ground state in this spatial domain (\fref{fig9abc_psiS395}a).

We also have to mention that for longer propagation times one can see appearing in the $|\Psi_{\Sigma}(k,t)|$ wavepacket peaks corresponding to the negative $k$ values of these other smaller momenta: -10.4, -8.6 and -6 a.u.

In the Appendix we use the impulsive approximation in the limit $\Delta(R) \gg W_L$  to show the emergence of such momenta during the time evolution. Indeed, as we emphasized, for the maximum of the ground wavefunction located at R $=89 \ a_0$, the impulsive approximation rests valid during the whole pulse (see \fref{fig9abc_psiS395}a), and for other maxima the impulsive approximation could be applied on smaller durations of the pulse. 

The significant fact is that these ``large momenta`` appearing at the end of the pulse in the ground state wavepacket are signatures of the maxima in the initial wavefunction continuum. This is a specific effect of the strong photoassociation pulse applied at small distances which reveals the structure of the  initial ground state.

\section{Comments and Conclusions} 
\label{sec:conclu}
We have analysed the dynamics in the photoassociation of a pair of cold atoms by a strong laser pulse (I $\approx$ 43 MW/cm$^2$) applied at short interatomic distances ($R_c=29.3 \ a_0$) for a cold collision. The numerical calculations were made for the $a^3\Sigma_u^+(6s+6s)$ $\to$ $1_g(6s+6p_{3/2})$ transition in Cs$_2$, at a temperature $T =$ 0.11 mK.
The large detuning ($\hbar\Delta_L=140$ cm$^{-1}$) and the intensity of the pulse were chosen
to correspond to a specific regime imposed by the limit of an
asymptotic detuning much bigger than the maximum of the coupling,
$\hbar\Delta_L=140$ cm$^{-1}$ $\gg W_L=13.17$ cm$^{-1}$, in order to avoid the transfer of population
 to the continuum of the excited state at the end of the pulse.

 The specificity of this
regime comes from two sides: a) the large detuning which locates the resonance
condition at small or intermediate interatomic distances, making
``visible''
the nodal structure of the initial continuum wavefunction; 
b) the strong coupling between the two electronic
channels, which acts not only on vibrational levels around the crossing of the dressed electronic
potentials, but also induces off-resonance cyclings of population between the coupled electronic states.

We have chosen not only a quite strong pulse, but also a quite long
one (a rectangular pulse of about 300 ps), in order to obtain a better
understanding of the time evolution in the presence of the pulse and to see how efficient for the cold molecule formation is the acceleration of the population from large to small interatomic distances. 

In this strong regime of coupling, the photoassociation dynamics during 
the pulse takes place in the light-induced (adiabatic) potentials, whose topology influence
the exchange of population between the coupled channels. In our example, the shapes of the adiabatic potentials 
lead to acceleration of the ground state population to the inner region (at short distances the diabatic potential $a^3\Sigma_{u}^{+}$ in $R^{-6}$ becomes the adiabatic one $V_{ad}^g$ decreasing in $R^{-3}$), and also
to the slowing down of the vibrational movement in the crossing region, which is
a mechanism for increasing progressively the population in the excited state. 

The main characteristics of the results at the end of the pulse are the following:

(i) It appears that such a pump scheme allows for the production of ground state molecules through a single laser pulse. Indeed, strongly bound cold molecules are formed in $a^3\Sigma_{u}^{+}$ and  $1_g$ electronic states, the
population transferred in bound levels of the ground state being much bigger than the population photoassociated in bound levels of the excited state. Some of the vibrational levels populated in $a^3\Sigma_{u}^{+}$ have  wavefunctions localized at distances $R < 35$ a$_0$. 
The final population in the $1_g$ excited state is entirely in bound vibrational levels, populated both
at resonance ($v_e=143$, with outer turning point at $\approx 30$ a$_0$) and off-resonance, where several
vibrational levels with outer turning points around $R \approx 70$
a$_0$  are populated due to the strong coupling catching the
large amplitude of the initial wavefunction between 45 and 65 a$_0$.
 This last kind of transfer creates a {\it hole} in the ground state in this domain of interatomic distances.

(ii) During the pulse the population in the ground state is globally accelerated towards small interatomic distances. The creation of a ``hole`` in the ground state wavepacket at the end of the pulse is also a factor leading to a compression of population at short distances \cite{elihole07}. 

(iii) At the end of the pulse, the momentum distribution of the ground
 state wavepacket keeps the traces of the initial continuum maxima. This is an effect of
the strong coupling leading to 
 off-resonance cycling of population between the two channels and bringing
 kynetic energy in the ground state. The cycling of population is particularly important
at those interatomic distances where the maxima of the initial continuum are located.

An important question is if the regime explored here can be mantained if, for example, one increases the detuning $\hbar\Delta_L$, in order to form cold molecules in lower vibrational levels of the ground state.  An insight about the results which can appear by increasing the detuning can be
extracted in {\it  the impulsive
limit}, which normally rests valid at large interatomic separations. 
In the impulsive limit, the population in
the excited state at the end of the pulse can be approximated as \cite{kosloff94}:
\begin{equation}
|\Psi_e(R,t)|^2 \approx \frac{W_L^2}{W_L^2 +\Delta^2(R)} \sin^2 (\Omega t) |\Psi_g(R,0)|^2,
\label{eq:peimp}
\end{equation}
where $\hbar\Omega(R)=\sqrt{W_L^2+\Delta^2(R)}$ (for $R \to \infty$, $2 \Delta(R) \to \hbar \Delta_L$). \Eref{eq:peimp} shows that the transfer from the ground to the excited state is favoured: a) at large interatomic distances R, because $|\Psi_g(R,0)|^2$ is the amplitude of an initial continuum wavefunction of low energy in the ground state; b) at resonance ($\Delta(R)=0$); c) for a strong coupling $W_L$, when, depending on the ratio $W_L/\hbar\Delta_L$ and the pulse duration, various ranges of distances can be
populated. It has to be observed that, increasing the detuning $\Delta(R)$, one
increases  $\Omega$, so the oscillations in $\sin^2 (\Omega t)$ with
the period $\pi/\Omega$ will be very fast and produce an unstable
regime of transfer, in which the result at the end of
the pulse cannot be predicted. In this regime, the
large distances  and the continuum of the excited state can easily rest
populated after the end of the pulse. The only manner to overcome these difficulties is to
increase the coupling, which can broke the impulsive
limit and bring population to short distances. Then, the ratio $W_L/\hbar\Delta_L$ is indeed a significant parameter for the regime explored here, which can be maintained for a larger detuning only if the pulse intensity is also increased.

The acceleration of the population to the inner region and the efficiency in forming strongly bound cold molecules is related here to a {\it non-impulsive regime of coupling} brought at $R < 70$ a$_0$ by the strength and the time duration of the pulse.

These results have to be completed with a further investigation of the time evolution subsequent to the pulse. Also the model can be enriched by considering couplings with other electronic surfaces. Our work is a tentative to show that the photoassociation of cold atoms at small/intermediate distances with an intense laser pulse offers the possibility to explore cold molecules formation using the specific topologies of the light-induced potentials. Favorable conditions can be found taking into account the variety of electronic transitions in alkali dimers which can be controlled by the parameters of the photoassociation pulse. 

\ack
Discussions with Prof. Ronnie Kosloff are gratefully acknowledged. 

\appendix
\section*{Appendix: Kinetic energy ``gains'' in the ground state due to the off-resonance cycling of population between the strongly coupled electronic states.}
\setcounter{section}{1}

\begin{figure}
\center
 \includegraphics[width=0.7\columnwidth]{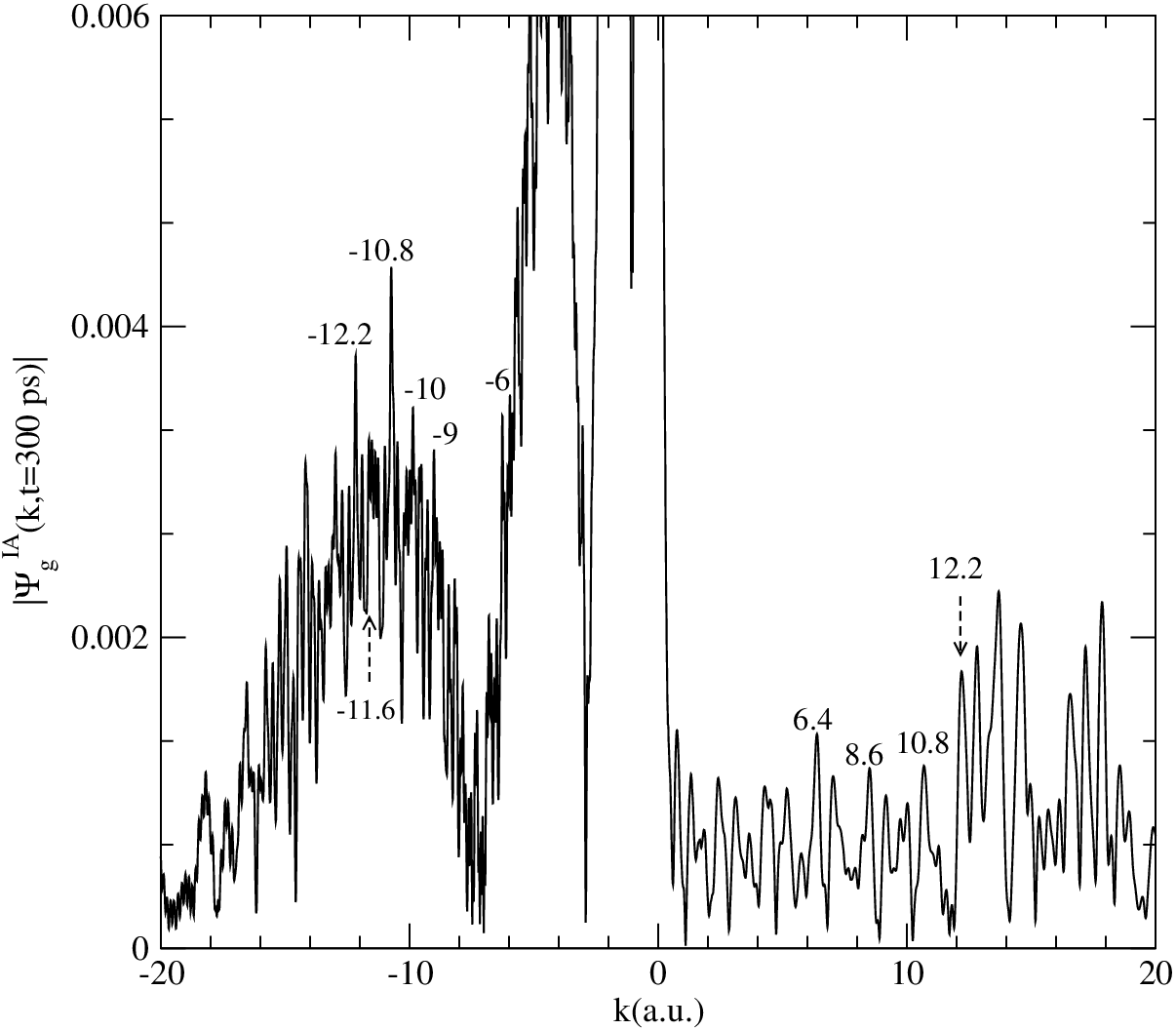}
\caption{The momentum representation $|\Psi^{IA}_{g}(k,t=300 \ ps)|$ of the wavefunction in the ground state $\Psi^{IA}_g(R,t)$, calculated in the impulsive approximation with \eref{eq:psigt}, for t=300 ps. The initial wavefunction $\Psi_g(R,0)$ has maxima at R = 33, 36.8, 41, 47, 59, 89 a$_0$ (see \fref{fig10_psir0max}), which, according to our reasoning, are supposed to give peaks in the momentum distribution $\Psi^{IA}_{g}(k,t)$ of the ground state, at $k \approx$ 6, 8.6, 10.4, 10.8, 11.6, 12.2 a.u. The figure shows that indeed the momentum distribution calculated in the impulsive approximation displays such peaks, at a time t close to the pulse duration. It can be noted that the amplitude of the $k=$-12.2 a.u. peak obtained here is close to 0.004, and that the same value can be observed in  \fref{fig9abc_psiS395}b) for  $|\Psi_{\Sigma}(k=\pm 12,2 \ a.u.,t=395 \ ps)|$.}
\label{fig11_psikIA300ps}
\end{figure}

We shall consider the time-dependent Schr\"odinger equation \eref{tschreq} for constant coupling $W_L$. In \cite{banin94} it is shown that, if the {\it impulsive
approximation} is valid at a given R, the wavefunction on the ground
state surface after a time $t$ becomes:
\begin{equation}
\Psi^{IA}_g(R,t)=\rme^{-\frac{\rmi}{\hbar} E_g t} \rme^{-\frac{\rmi}{\hbar} \Delta t}
\{ \cos(\Omega t) + \rmi \cos \theta \sin(\Omega t) \} \Psi_g(R,0),
\label{eq:psigt}
\end{equation}
where $E_g$ is the energy corresponding to the initial stationnary
wavefunction at t=0 ($[T+V_g]\Psi_g(R,0)=E_g \Psi_g(R,0)$),
2$\Delta(R)=|V_e(R)-V_g(R)|$ is the local detuning, $\Omega(R)$ the
spatial dependent Rabi pulsation:
\begin{equation}
\hbar \Omega (R) = {\sqrt{W_L^2+\Delta(R)^2}} 
\end{equation}
and 
\begin{equation}
\cos\theta(R)= \frac{\Delta(R)}{\hbar \Omega(R)}= \frac{\Delta(R)}{\sqrt{W_L^2+\Delta(R)^2}}.
\end{equation}

 We calculated numerically the time-dependent prediction for the ground state wavefunction in the impulsive approximation, $\Psi^{IA}_g(R,t)$, using \eref{eq:psigt}. 
In \fref{fig11_psikIA300ps} we show the corresponding momentum distribution at t=300 ps, $|\Psi^{IA}_{g}(k,t=300 \ ps)|$, which indeed displays peaks at $k$ values deduced from the maxima of the initial wavefunction, according to the discussion of \Sref{sec:kfeatures}. 

 In the following  we shall introduce approximations into \eref{eq:psigt} in order to make appear
analitically a momentum $k$ associated with a maximum of the initial wavefunction at $R_0$.

In (\ref{eq:psigt}), $\rme^{-\frac{\rmi}{\hbar} E_g t}
\Psi_g(R,0)$ represents a free evolution in the ground
state. The factor superposed on this free evolution can be separated:
\begin{eqnarray}
 A(R,t) &= \rme^{-\frac{\rmi}{\hbar} \Delta t}
\{ \cos(\Omega t) + \rmi \cos \theta \sin(\Omega t) \} \\
        & =\frac{1}{2} \rme^{\rmi(\Omega-\frac{\Delta}{\hbar})t}
  (1+\frac{\Delta}{\hbar\Omega})+
\frac{1}{2} \rme^{-\rmi(\Omega+\frac{\Delta}{\hbar})t}
  (1-\frac{\Delta}{\hbar\Omega}) \label{at}
\end{eqnarray}
We shall make approximations on the expression (\ref{at}) in 
 the limit \underline {$\Delta(R) \gg
  W_L$}. Indeed, this limit is valid at $R_0=89 \ a_0$, where
$\Delta(R_{0})=\hbar\Delta_L/2 \approx 70$ cm$^{-1}$, and
$W_L=13.17$ cm$^{-1}$. Then it is possible to approximate:
\begin{eqnarray}
\hbar \Omega= \Delta \sqrt{1+ \frac{W^2}{\Delta^2} } \approx \Delta 
+ \frac{W^2}{2\Delta}.
\end{eqnarray}
Consequently, at a given R value where the impulsive approximation is valid,
$\Psi_g(R,t)$ can be decomposed in two terms:
\begin{eqnarray}
\Psi^{IA}_g(R,t) \approx \Psi^{(1)}_g(R,t) +\Psi^{(2)}_g(R,t) \\
\Psi^{(1)}_g(R,t) = \frac{1}{2}(1+\frac{\Delta}{\hbar\Omega})
                  \rme^{ \frac{\rmi}{\hbar} \frac{W^2}{2\Delta} t } 
                  \rme^{-\frac{\rmi}{\hbar} E_g t}  \Psi_g(R,0) \\
\Psi^{(2)}_g(R,t) = \frac{W_L^2}{4 (\hbar\Omega) \Delta}
                  \rme^{ - \frac{\rmi}{\hbar}[2\Delta+ \frac{W^2}{2\Delta}] t } 
                  \rme^{-\frac{\rmi}{\hbar} E_g t}  \Psi_g(R,0)
\end{eqnarray}

For $\Delta \gg W_L$ the first term is the
dominant one. Indeed, a rough approximation $\hbar\Omega \approx
\Delta$ implies $|\Psi^{(1)}_g(R,t)| \approx
|\Psi_g(R,0)|$, the second term having an incomparable smaller
contribution ($\frac{W_L^2}{4 (\hbar\Omega) \Delta}|_{89 a_0}=0.0087 \ll 1$). It is the 
second ``small'' term which offers the explanation for the ``k-
features'' observed in our results.  Its Fourier transform  is:
\begin{eqnarray}
\Psi^{(2)}_g(k,t) = \int dR e^{-\rmi k R} \Psi^{(2)}_g(R,t)
\\
= \rme^{-\frac{\rmi}{\hbar} E_g t} \int dR \frac{W_L^2}{4 (\hbar\Omega) \Delta}
                  e^{ -\rmi[ \frac{E_{2\Delta}}{\hbar} t  +kR ] }   \Psi_g(R,0)
\end{eqnarray}
with $E_{2\Delta}= 2\Delta+ \frac{W^2}{2\Delta} \approx 2\Delta $.
We shall take into account only the contribution at the integral
coming from a small domain of R, $D_R$, around
the point $R_0$, assuming that on this domain $\Psi_g(R,0)$ is a
Gaussian of width $\delta_R$ centered in $R_0$: $\Psi_g(R,0)|_{R_0} \approx \Psi_g(R_0,0) \exp[-(R-R_0)^2/\delta_R^2]$, and that $E_{2\Delta}(R)$ depends linearly on R near $R_0$: 
 $E_{2\Delta}(R) \approx $ $E_{2\Delta}(R_0)+ \frac{dE_{2\Delta}}{dR}|_{R_0}(R-R_0)$. 
Then:
\begin{eqnarray}
\Psi^{(2)}_g(k,t)|_{R_0} &\approx \rme^{-\frac{\rmi}{\hbar} E_g t}  \frac{W_L^2}{4 \hbar\Omega(R_0) \Delta(R_0)} \Psi_g(R_0,0) \rme^{ -\rmi[\frac{E_{2\Delta}(R_0)}{\hbar} t  +kR_0]} \nonumber\\
                     &\times  \int_{-\delta_R}^{\delta_R}dr  \rme^{ -\rmi[\frac{1}{\hbar}
               \frac{dE_{2\Delta}}{dR}|_{R_0} t  +k]r}  \rme^{-r^2/\delta_R^2}
\label{psik2}
\end{eqnarray}
where $r=R-R_0$ and the initial  domain of integration was taken from $R_0-\delta_R$ to $R_0+\delta_R$.
For a $\delta_R$ sufficiently large, the integral in \eref{psik2} becomes a $\delta$ function, such as:
\begin{eqnarray}
\Psi^{(2)}_g(k,t)|_{R_0} &\approx \rme^{-\frac{\rmi}{\hbar} E_g t}  \frac{W_L^2}{4 \hbar\Omega(R_0) \Delta(R_0)} \Psi_g(R_0,0) \rme^{ -\rmi[\frac{E_{2\Delta}(R_0)}{\hbar} t  +kR_0]} \nonumber\\
                     &\times 2 \pi \delta (\frac{1}{\hbar}\frac{dE_{2\Delta}}{dR}|_{R_0} t + k )
\label{psik2delta}
\end{eqnarray}
In \eref{psik2delta} one can see appearing the
contribution of an ingoing ($kR_0=-\vec{k} \vec{R_0}$) plane wave $\rme^{ -\rmi[ \frac{E_{2\Delta}(R_0)}{\hbar} t  +kR_0]}$,
  of energy $E_{2\Delta}(R_0)/\hbar$ and momentum $p=\hbar k$ \cite{messiah}. The $\delta$ function shows that this momentum $p=\hbar k$ = $-\frac{dE_{2\Delta}}{dR}|_{R_0} t$ is related to the local energy difference between the electronic potentials at $R_0$, or that $dp/dt = -\frac{dE_{2\Delta}}{dR}|_{R_0}$. If one associates this momentum to the movement of an ingoing particle of mass $\mu$,  $p=-\mu dR/dt$,  then one obtains $p/\mu=dE_{2\Delta}/dp$, and finally $E_{2\Delta}=p^2/2\mu$. 

Then indeed \eref{psik2delta} makes appear a momentum $k$ connected to $2\Delta(R_0)$, as we observed in our results discussed in \Sref{sec:kfeatures}, and oriented to small R distances. The fact that we first observe an outgoing wave has to be due to the rapid reflection of the ingoing waves by the inner wall of the ground
state potential.

 An estimation of the wavefunction amplitude $\Psi^{(2)}_g(k,t)|_{R_0}$ for $R_0=89\ a_0$ using formula \eref{psik2delta} gives $|\Psi^{(2)}_g(k,t)|_{R_0=89 a_0}  \approx 0.001 \times \delta (\frac{1}{\hbar}\frac{dE_{2\Delta}}{dR}|_{R_0} t + k )$. This qualitative estimation obtained analitically for unprecised $t$  corresponds to the amplitudes calculated numerically for $k= \pm$ 12.2 a.u., from the dynamics and in the impulsive approximation, which are in agreement: $|\Psi_{\Sigma}(k= \pm 12,2 \ a.u.,t=395 \ ps)|$ $\approx$ $|\Psi^{IA}_{g}(k=-12.2 \ a.u.,t=300 \ ps)|$ $\approx$ 0.004 (see \fref{fig9abc_psiS395}b) and \fref{fig11_psikIA300ps}).

\end{document}